\documentclass[journal]{IEEEtran}
\usepackage{algorithm}
\usepackage{algpseudocode}
\usepackage{multirow}
\usepackage[table ]{xcolor}
\usepackage{array}
\usepackage{algpseudocode}
\usepackage{amsmath,amssymb}
\usepackage{cite}
\usepackage{booktabs}
\usepackage{graphicx}
\usepackage{float}
\usepackage{threeparttable}
\usepackage{epstopdf}
\usepackage{xfrac}
\usepackage{makecell}
\usepackage{cleveref}
\usepackage{setspace}
\usepackage{graphicx}
\usepackage{mathtools, cuted}
\usepackage[english]{babel}
\usepackage{lipsum}
\usepackage{subfigure}
\usepackage{caption}
\usepackage{url}

\begin{document}
\title{Defending Adversarial Examples by Negative Correlation Ensemble}

\author{Wenjian~Luo,~\IEEEmembership{Senior Member,~IEEE}, Hongwei~Zhang, Linghao~Kong, Zhijian Chen, Ke~Tang,~\IEEEmembership{Senior Member,~IEEE}
	\thanks{This study is supported by the Major Key Project of PCL (Grant No. PCL2022A03, PCL2021A02, PCL2021A09), and the Guangdong Provincial Key Laboratory (Grant No. 2020B121201001). \textit{(Corresponding author: Wenjian Luo.)}}
	\thanks{Wenjian Luo, Hongwei Zhang, Linghao Kong and Zhijian Chen are with the School of Computer Science and Technology, Harbin Institute of Technology, Shenzhen 518055, Guangdong, China. Wenjian Luo is also with Peng Cheng Laboratory, Shenzhen 518055, China.

	Ke Tang is with Guangdong Provincial Key Laboratory of Brain-inspired Intelligent Computation, the School of Computer Science and Engineering, Southern University of Science and Technology, Shenzhen 518055, Guangdong, China.
	}
	
	\thanks{
	Email: 
	luowenjian@hit.edu.cn, 
	20S151127@stu.hit.edu.cn, 
	(20S151073, 21B951010)@stu.hit.edu.cn, 
	tangk3@sustech.edu.cn.
	}
}

\maketitle

\begin{abstract}
The security issues in DNNs, such as adversarial examples, have attracted much attention. 
Adversarial examples refer to the examples which are capable to induce the DNNs return completely predictions by introducing carefully designed perturbations.
Obviously, adversarial examples bring great security risks to the development of deep learning. 
Recently, Some defense approaches against adversarial examples have been proposed, 
however, in our opinion, the performance of these approaches are still limited.
In this paper, we propose a new ensemble defense approach named the Negative Correlation Ensemble (NCEn), 
which achieves compelling results by introducing gradient directions and gradient magnitudes of each member in the ensemble negatively correlated and 
at the same time, reducing the transferability of adversarial examples among them.
Extensive experiments have been conducted, and the results demonstrate that NCEn can improve the adversarial robustness of ensembles effectively.
\end{abstract}


\section{Introduction}
\label{introduction}
Deep Neural Networks (DNN) have achieved significant improvements in various domains,
typically like image classification, face recognition and autonomous driving \cite{Lecun2015, DBLP:conf/nips/KrizhevskySH12, DBLP:conf/aaai/YanXL18, Sun2014, Akhtar2020, 7995703}. 
However, there are serious security issues in DNNs which have attracted much attention in recent years \cite{DBLP:journals/corr/SzegedyZSBEGF13}. 
Specifically, studies have shown that DNN are vulnerable to the attacks from adversarial examples, 
which are generated by adding delicate and imperceptible perturbation to the benign samples, 
and aims to prompt the DNNs to make incorrect predictions\cite{Goodfellow2015}. 

Many algorithms have been proposed for generating adversarial examples \cite{Goodfellow2015, Kurakin2019, DBLP:conf/cvpr/Moosavi-Dezfooli16, DBLP:conf/sp/Carlini017, DBLP:conf/iclr/BrendelRB18,
DBLP:conf/sp/ChenJW20, dong2018boosting, DBLP:conf/cvpr/XieZZBWRY19, DBLP:conf/iclr/Wu0X0M20}. 
These algorithms can be divided into white-box attacks and black-box attacks according to the accessibility.
White-box attack algorithms, such as Fast Gradient Sign Method (FGSM) \cite{Goodfellow2015}, 
Iterative Fast Gradient Sign Method (I-FGSM) \cite{Kurakin2019},
DeepFool \cite{DBLP:conf/cvpr/Moosavi-Dezfooli16} and C$\&$W \cite{DBLP:conf/sp/Carlini017}, 
exploit both the network structures, hyperparameters and other available information of the target model, 
to calculate the gradient and such that generate adversarial examples. 
On the contrary, black-box attack algorithms, 
such as Zero Order Optimization (ZOO) \cite{DBLP:conf/ccs/ChenZSYH17}, 
Autoencoder based Zero Order Optimization (AutoZoom) \cite{DBLP:conf/aaai/TuTC0ZYHC19}, 
Boundary Attack \cite{DBLP:conf/iclr/BrendelRB18}, HopJumpSkippAttack\cite{DBLP:conf/sp/ChenJW20},  
Momentum Iterative Fast Gradient Sign Method (MI-FGSM) \cite{dong2018boosting}, 
Diverse Input Iterative Fast Gradient Sign Method (DIM) \cite{DBLP:conf/cvpr/XieZZBWRY19} and 
Skip Gradient Method (SGM) \cite{DBLP:conf/iclr/Wu0X0M20}, 
are capable to generate adversarial examples without accessing the all the information of the target model (e.g. not accessing the target model 
but leveraging a substitution model). 
Notably, the black-box algorithms such as MI-FGSM can also be used in white-box setting scenarios, where they can directly access the target model. 

Effective defenses against adversarial examples are usually achieved by detecting adversarial perturbations 
or improving the robustness of the model. 
The approaches of detecting adversarial perturbations, 
such as the key-based network in \cite{DBLP:journals/corr/abs-1806-00580}, the MagNet framework in \cite{DBLP:conf/ccs/MengC17}
and the feature squeezing in \cite{DBLP:conf/ndss/Xu0Q18}, 
are implemented mainly by detecting or cleaning the input data through technical means, 
so that can discover the adversarial examples in advance or destroy some key structures constituting the adversarial examples. 
Differently, the approaches of improving the robustness of the model, 
such as the gradient regularization \cite{DBLP:journals/corr/abs-1806-08028}, adversarial training \cite{kurakin2016adversarial}  
and the defensive distillation \cite{DBLP:conf/sp/Carlini017}, 
are implemented by changing some specific properties of the target model, 
such that enhance the robustness of the model to small perturbations.

Recently, some studies have found that, for a same learning task, different models will learn different decision boundaries 
due to the differences in their model structures, initial weights, and training methods, 
and by which we can infer that an adversarial example which can fool one model may not be capable to fool another ones \cite{DBLP:journals/corr/abs-1712-04006}.
Intuitively, the vulnerability of a single model can be avoided through model ensemble. 
Kariyappa et al. \cite{2019Improving} have proposed the Gradient Alignment Loss (GAL) to improve the diversity of ensemble by reducing the dimension of shared adversarial subspace  (ADV-SS). 
Pang et al. \cite{DBLP:conf/icml/PangXDCZ19} have proposed a training method named Adaptive Diversity Promoting (ADP) to 
make the non-maximum predictions of different members in the ensemble orthogonal to each other, 
so that can improve the diversity of the ensemble without compromising the prediction accuracy. 
What's more, considering the influence of the gradient magnitude and the gradient angle in an ensemble,
Dabouei et al. \cite{2020Exploiting} have proposed joint Gradient Phase and Magnitude Regularization (GPMR) based on GAL 
to obtain the effective defense interaction 
by considering both the optimal geometric boundary and the gradient magnitude of the members in the ensemble.
However, although the ensemble methods can achieve adversarial robustness by increasing the diversity, 
the complementarity among members is neglected during the training which may lead to low Performance when facing various adversarial examples attack. 
For example, an adversarial example which can fool one member of the ensemble may also be capable to fool the other members due to its transferability.

In this paper, in order to leverage the interactions between members in the ensemble and reduce the transferability of adversarial examples among members, 
we propose to train classifiers in the ensemble based on the negative correlation principle and accordingly, design a new ensemble defense strategy (NCEn) to 
improve the adversarial robustness of ensembles. 
In NCEn, we make the gradient angle and the gradient magnitude of each member with respect to the input $x$ negatively correlated, 
thus prompting the gradient directions to have the greatest diversity, and the gradient magnitude to be balanced. 
Specifically, for realizing the greatest diversity of gradient directions based on negative correlation, 
the cosine similarity between the gradients of any member and the ensemble is trained negatively correlated with the cosine similarity 
between the ensemble and all other members. 
So that we can reduce the number of members that are vulnerable to adversarial perturbation $\epsilon$. 
After this, we make the gradient magnitude of any member in the ensemble to be negatively correlated with the gradient magnitude of other members in the ensemble,
which indicates that the prediction results of the ensemble will not be greatly compromised by the maximum gradient magnitude of individual members in the ensemble.

The contributions of this paper can be summarized as follows.
\begin{itemize}
	\item  We propose a novel negative correlation ensemble (NCEn).
	Based on the principle of negative correlation, 
	NCEn maximizes the diversity of members in the ensemble and reduces the transferability of adversarial examples among members in the ensemble 
	by constraining the gradient direction and the gradient magnitude of each member. 
	\item  Extensive experiments have been conducted, and the results show that the performance of 
	the negative correlation ensemble (NCEn) exceeds the state-of-the-art ensemble-based defense strategies.
	What's more, the experimental results also show that NCEn can effectively reduce the transferability of adversarial examples 
	as well as improving the diversity and robustness of the ensemble.
\end{itemize}

The rest of this paper is organized as follows. 
Section \ref{sec: related work} introduces related work. 
We then introduce our approach in Section \ref{sec: Adversarial Robustness of Ensemble}. 
Section \ref{sec:Experiment} presents our experimental details and results. 
A brief conclusion of this paper is given in Section \ref{sec: Conclusion}.

\section{Related Work} 
\label{sec: related work}
In this section, we first introduce the concept of adversarial examples as well as the generation and defense algorithms. 
After this, we will introduce the adversarial robustness of ensembles and negative correlation learning, respectively.

\subsection{Adversarial Examples}
\label{sec:Adversarial Examples}
In 2014, Szegedy et al. \cite{DBLP:journals/corr/SzegedyZSBEGF13} have proposed the concept of adversarial examples. 
Given a trained classifier $f(\cdot)$ and a clean input sample $x$ with the corresponding label $y$, 
the adversarial example $x' = x + \epsilon$ satisfies $f(x') \neq y$, where $\epsilon$ is a subtle perturbation that is not easily detected by human eyes.

Numerous adversarial examples generation algorithms have been proposed in past years
\cite{Goodfellow2015, dong2018boosting, DBLP:conf/iclr/MadryMSTV18, Kurakin2019,DBLP:journals/corr/SzegedyZSBEGF13, DBLP:conf/sp/Carlini017, DBLP:conf/cvpr/Moosavi-Dezfooli17, DBLP:conf/cvpr/Moosavi-Dezfooli16}.
Goodfellow et al. \cite{Goodfellow2015} have proposed the FGSM algorithm based on the linear hypothesis of the high-dimensional space of deep neural networks. 
The advantage of FGSM is that it can generate adversarial examples quickly, and can be applied to a variety of deep neural networks. 
The momentum iterative FGSM (MI-FGSM) \cite{dong2018boosting} is an iterative attack method based on FGSM. 
By adding a momentum term to the iterative attack method, the transferability of adversarial examples is improved.
The project Gradient Descent (PGD) \cite{DBLP:conf/iclr/MadryMSTV18} is another iterative attack method based on FGSM. 
In each iteration process, the perturbation is clipped to the specified range to generate efficient adversarial examples. 
In 2017, Kurakin et al. \cite{Kurakin2019} have proposed the basic iterative method (BIM). 
By limiting each pixel of the adversarial example to be within the $l_p$ field of the original example $x$, 
they constructed the adversarial examples in the real scenarios. 
Compared to FGSM, BIM makes fewer changes to the original sample and has better attack Performance.

In order to defend against the adversarial examples,
a lot of methods have been proposed, such as adversarial training, networks modification, feature squeezing as well as using additional networks \cite{DBLP:journals/corr/abs-1806-00580,DBLP:journals/corr/abs-1806-08028,
kurakin2016adversarial,DBLP:conf/nips/GoodfellowPMXWOCB14,DBLP:conf/sp/Carlini017,
DBLP:conf/sp/PapernotM0JS16, DBLP:conf/iclr/SamangoueiKC18, DBLP:conf/ccs/MengC17, DBLP:conf/ndss/Xu0Q18, DBLP:journals/corr/DziugaiteGR16, 
DBLP:journals/tnn/YuanHZL19, DBLP:conf/nips/ZhengH18}. 
Specifically, the method of adversarial training improves the robustness of the model by generating adversarial examples and updating model parameters 
alternatively \cite{DBLP:journals/corr/SzegedyZSBEGF13, Goodfellow2015, DBLP:conf/cvpr/Moosavi-Dezfooli16, Kurakin2019, DBLP:conf/iclr/MadryMSTV18}. 
In order to keep the effectiveness, this method needs to use high-intensity adversarial examples, 
and at the same time, the network should have sufficient expressive power. 
Therefore, such method requires a large amount of adversarial example training data, so it is also called brute force adversarial training. 
In addition, existing works show that adversarial training can help regularize the models and mitigate overfitting \cite{Lecun2015, 
sankaranarayanan2018regularizing23}. 
However, these adversarial trained models still lack robustness to unseen adversarial examples.

\subsection{Adversarial Robustness of Ensembles}
\label{sec:Adversarial Robustness of Ensembles}
An efficient way to defend against adversarial examples is to use an ensemble of deep neural networks \cite{2019Improving, 
DBLP:conf/iclr/AbbasiG17, DBLP:journals/corr/abs-1712-04006, DBLP:conf/icml/PangXDCZ19}. 
By using different model structures, initial weights, and training methods for different models can significantly improve the diversity. 
Strauss et al. \cite{strauss2017ensemble} have shown that neural networks the ensemble can not only improve the prediction accuracy, 
but also improve the robustness to adversarial examples. 
What's more, 
Tramèr et al. \cite{tramer2017ensemble} have proposed to use the adversarial training to enhance the adversarial robustness of an ensemble.

Since the gradients of the models in the ensemble share similar directions, 
an adversarial example which can fool one model may also be capable to fool the others in the ensemble. 
For this, Kariyappa et al. \cite{2019Improving} have proposed Gradient Alignment Loss (GAL) to improve the adversarial robustness of the ensemble 
by considering the diversity of gradient directions. 
Specifically, it focuses on training the ensemble whose members have irrelevant loss functions by using diversified training. 
Intuitively, GAL reduces the dimension of adversarial subspace of the models through diversified training. 
However, GAL does not consider the optimal geometric boundary for the diversification of gradient directions in the ensemble, 
and does not balance the gradient magnitude of each member in the ensemble. 
Based on this, Dabouei et al. \cite{2020Exploiting} have proposed gradient phase and magnitude regularization (GPMR) \cite{2020Exploiting}. 
The basic principle of GPMR is to increase the lower bound of the adversarial perturbation that changes the score of the classifier by considering the 
optimal geometric boundary to diversify the gradient direction in the ensemble, and to balance the gradient magnitude of the members, 
thereby constructing the first-order defense interaction of the members in the ensemble. 
However, GPMR does not fully consider the interaction between members in the ensemble. 
In contrast, our work builds a good ensemble defense system by considering the interaction between members in the ensemble, thereby improving the robustness of the ensemble.

\subsection{Negative Correlation Learning}
\label{sec:Negative Correlation} 
Negative correlation ensemble has been proposed by Liu and Yao in 1999 \cite{liu1999simultaneous, DBLP:journals/nn/LiuY99}. 
Liu and Yao \cite{liu1999simultaneous} have proposed a cooperative ensemble learning system (CELS), 
and the purpose is to interactively and simultaneously train all individual models in the ensemble in a single learning process 
through negative correlation learning (NCL). 
Through negative correlation learning, different models in the ensemble can learn different features of the training set, 
so that the ensemble can work more holistically. 
Theoretical and experimental results show that NCL can promote the diversity of the models as well as keeping a high prediction accuracy. 

Liu et al. \cite{DBLP:journals/tec/LiuYH00} also have tried to solve the problem of the optimal number of neural networks in the ensemble on the premise of maintaining the good interaction 
of individual members in the ensemble through negative correlation learning and evolutionary learning. 
Chan et al. \cite{DBLP:journals/npl/ChanK05} proposed NCCD, which implemented negative correlation learning via correlation corrected data. 
NCCD does not add penalty items in the training process, but adds error related information to the training data for negative correlation learning. 
NCCD can reduce the communication bandwidth between individual networks, and can be applied to the ensemble of any type of network structures. 
In addition, because NCCD does not modify the error function of the network for negative correlation learning, but modifies the training data for negative correlation learning, 
it can accelerate the learning speed through parallel computing. 

Wang et al. \cite{DBLP:conf/ijcnn/WangCY10} proposed a new negative correlation learning algorithm AdaBoost.NC. 
The flexibility of AdaBoost is used to overcome the shortcomings of NCL, such as sensitivity to parameters setting and long training time, 
and the overfitting problem of AdaBoost is solved by introducing diversity \cite{DBLP:conf/aaai/Quinlan96, DBLP:journals/jair/OpitzM99}. 
The experimental results show that AdaBoost.NC has better generalization performance than NCCD, and the time cost is significantly less than CELS and NCCD.

In this paper, for the first time, we propose a defense approach against adversarial examples by leveraging the negative correlation method. 
We first make the gradient direction of each member in the ensemble negatively correlated. 
Secondly, we make the gradient magnitude of each member in the ensemble negatively correlated with the average gradient magnitude, 
such that help improve the diversity of the ensemble and balance the gradient magnitude of each member in the ensemble. 
The details will be given in the next section.

\section{The Proposed Method}
\label{sec: Adversarial Robustness of Ensemble}
The purpose of defense is to enable the classifier $f$ give a correct prediction on examples $x'$ which is with the adversarial perturbation. 
The ensemble strategy exploits multiple trained models to make decisions together. 
Due to the differences in structures, initial weights and training methods, different models could learn different decision boundaries. 
This means that an adversarial example which can fool one model may not be capable to fool the others. 
So the ensemble could have better adversarial robustness compared to a single model.

In this section, we will show how negative correlation theory can be utilized in ensemble training and how it can be exploited to improve the adversarial robustness.

\subsection{On Gradient Directions of Members}
\label{sec:Use negative correlation to increase ensemble diversity}
The gradient of the loss to the input $x$ refers to a direction where the directional derivative of the loss achieves a maximum along this direction, 
i.e., the loss changes fastest when a perturbation $\epsilon$ is introduced along this direction.  
When the members in the ensemble have different gradient directions, 
we put the adversarial example $x'$ with perturbation $\epsilon$ along a certain direction into the ensemble for prediction, 
and $x'$ could significantly change only the loss function of classifiers with gradient direction similar to $\epsilon $, but not affect others. 

We treat the mean value of the gradients of all members as the gradient of the ensemble, 
which can be represented by $\nabla J_{ensemble}$.  
The adversarial perturbation added along the direction of $\nabla J_{ensemble}$ can affect most members of the ensemble. 
We use $\nabla J_i$ to represent the gradient of the $i^{th}$ member with respect to input $x$. 
Firstly, we prompt the angle between $\nabla J_i$ and $\nabla J_{ensemble}$ negatively correlated with the angle 
between $\nabla J_j$ ($j \neq i$) of other members and $\nabla J_{ensemble}$. 
After negative correlation training, the gradient direction of all members in the ensemble with respect to whole dataset will be maximally different, 
which means that an adversarial perturbation increasing the loss function of one member will not absolutely increase the loss of other members.

In detail, we use cosine similarity (CS) to measure the gradient angle between members, 
and it can be calculated as Formula \eqref{CS}. 

\begin{equation}\label{CS}
	CS(\nabla J_i, \nabla J_j) = \frac{<\nabla J_i, \nabla J_j>}{\|\nabla J_i\|_2 \cdot \|\nabla J_j\|_2} 
\end{equation}
 
The smaller the CS, the larger the angle between the gradients, and the greater the diversity among the members. 
This means that the loss of each member will not grow in a positive correlation manner for a same adversarial perturbation. 
Therefore, it is hard for an adversarial perturbation to fool all members simultaneously. 
Meanwhile, the ensemble with increased diversity is still capable to keep a high prediction accuracy. 
The relevant regularization term can be expressed as Formula \eqref{negative correlation diversity}.

\begin{equation}\label{negative correlation diversity}
	\operatorname{Loss}_{\mathrm{cos}}=\operatorname{CS}\left(\nabla J_i, \nabla J_{ensemble}\right) \sum_{j \neq i}^{\text{k}} \operatorname{CS}\left(\nabla J_j, \nabla J_{ensemble}\right)
 \end{equation}

\begin{equation}
	\nabla J_{ensemble}=\frac{1}{k} \sum_{i=1}^{\text{k}}\nabla J_i 
\end{equation} 

Where $\nabla J_{ensemble}$ is the averaged gradients of all members, $k$ is the number of members, 
and $\nabla J_i$ is the gradient of the $i^{th}$ member.

\begin{figure}[h]
	\centering
	\includegraphics[width=.4\textwidth]{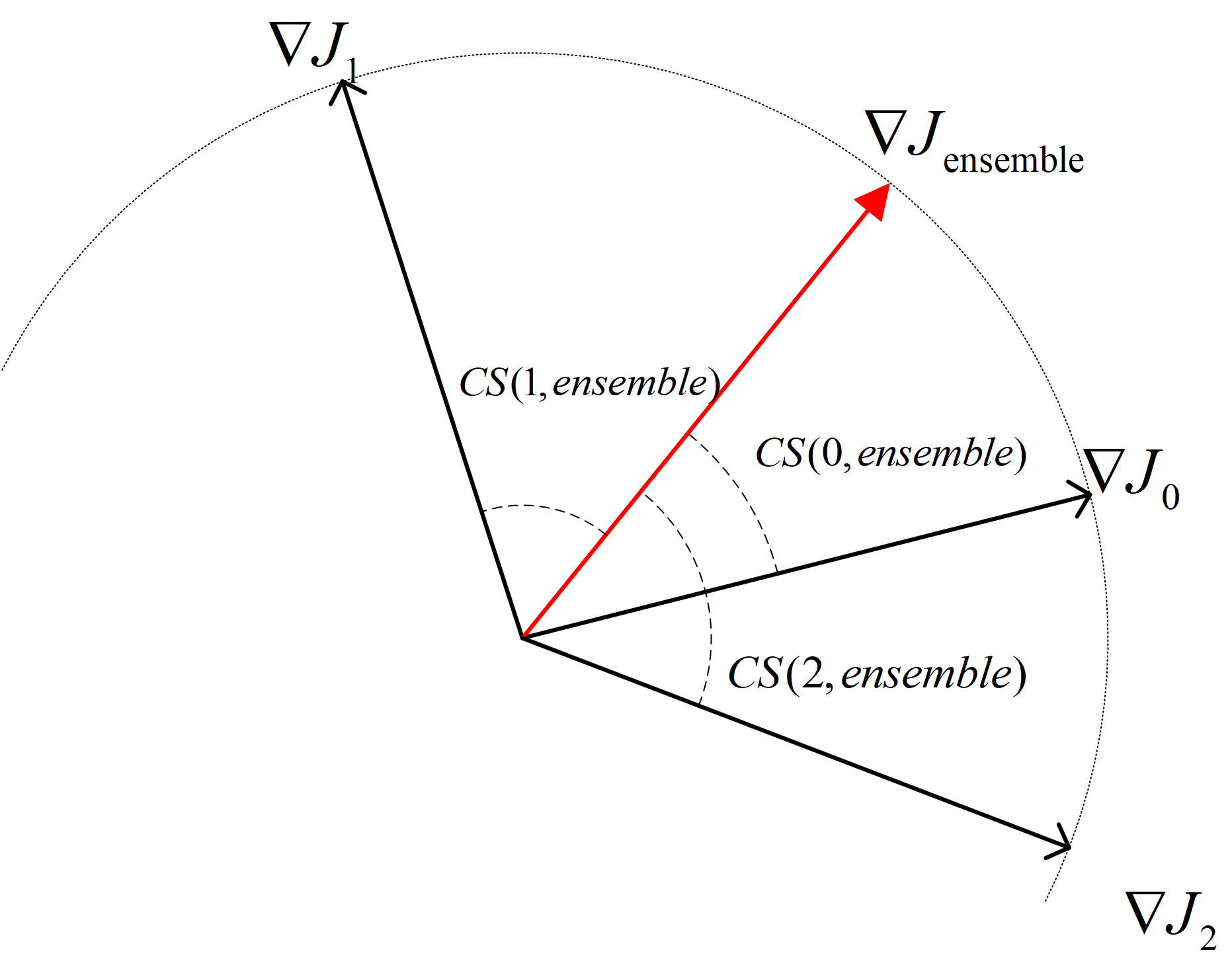} 
	\caption{Improve the ensemble diversity from the view of gradient direction; $\nabla J_i$ indicate the model gradient of the $i^{th}$ member ($i=0,1,2$), and $\nabla J_{ensemble}$ indicate the ensemble gradient} 
	\label{diversity} 
\end{figure}

In the training process, 
the regularization term aims to prompt the gradient direction of each member in the ensemble negatively correlated with the others. 
From Fig. \ref{diversity}, where $\nabla J_i$ indicates the model gradient of the $i^{th}$ member($i=0,1,2$), 
and $\nabla J_{ensemble}$ indicates averaged gradients of all members, 
the follows can be observed:

\begin{itemize}
	\item When $\sum_{j \neq i}^k CS\left(\nabla J_j, \nabla J_{ensemble}\right)$ is greater than zero, the gradient directions of $\nabla J_j$ are closer to $\nabla J_{ensemble}$. 
	In such a situation, minimizing the negative correlation formula (\ref{negative correlation diversity}) will result in a larger gradient angle between $\nabla J_i$ and $\nabla J_{ensemble}$.
	\item When $\sum_{j \neq i}^k CS\left(\nabla J_j, \nabla J_{ensemble}\right)$ is negative, the gradient directions of $\nabla J_j$ is inconsistent with that of $\nabla J_{ensemble}$. 
	In such a situation, minimizing the negative correlation formula (\ref{negative correlation diversity}) will result in the gradient direction of $\nabla J_i$ to approach the gradient direction of $\nabla J_{ensemble}$.
\end{itemize}

In both cases, the regularization term will help all members train interactively, 
such that facilitates the maximization of the gradient direction diversity.

We use CS as the regularization term for enhancing the adversarial robustness, 
and the loss function can be expressed as Formula \eqref{loss diversity},
where $CE$ is the mean of the cross-entropy loss and $\lambda_{cos}$ is the weight coefficient.

\begin{equation}\label{loss diversity}
	Loss=CE+\lambda_{cos} Loss_{cos}
\end{equation}

\begin{equation}\label{CE}
	CE= \frac{1}{k} \sum_{i=1}^k ce_{loss}
\end{equation}

\begin{algorithm}[!ht]
	\caption{Calculating $Loss_{cos}$ for the $i^{th}$ member}
	\label{alg: diversity}
	\begin{algorithmic}
		\Require
		Gradient of the $i^{th}$ member: $\nabla J_i$;
		Gradient of other members in the ensemble: $\nabla J_j(j \neq i)$;
		Ensemble gradient: $\nabla J_{ensemble}$;
		Number of models in the ensemble: $k$
		\Ensure
		Regularization term of the $i^{th}$ member: $Loss_{cos}$
		
		\State $cos1 \gets CS(\nabla J_i, \; \nabla J_{ensemble})$
		\State $cos2 \gets 0$;
		\For{$j \gets 1 \; \textbf{to} \; k$}
				\If {$j \neq i$}
					\State $cos2 \gets cos2+CS(\nabla J_j, \; \nabla J_{ensemble})$
				\EndIf
		\EndFor
			\State $Loss_{cos} \gets cos1 * cos2$			 
		\State Return the regularization term: $Loss_{cos}$
 
	\end{algorithmic}
\end{algorithm}

The algorithm for calculating the regularization term of the $i^{th}$ member on the gradient direction using negative correlation is shown in Algorithm \ref{alg: diversity}, 
where $k$ is the number of members in the ensemble. 
Firstly, it calculates the cosine similarity $cos1$ between the gradient of the $i^{th}$ member and the ensemble as well as 
the sum of cosine similarity $cos2$ between the gradients of the other members and the ensemble. 
After this, the negative correlation coefficient of $cos1$ and $cos2$ are calculated 
as the regularization term $Loss_{cos}$ of the $i^{th}$ member and will be minimized in the training process.

\subsection{On the Gradient Magnitudes of Members}
\label{sec:Use negative correlation to equalize the ensemble gradient}
The gradient magnitude represents the magnitude of the change in the loss caused by the adversarial perturbation $\epsilon$. 
Adding an adversarial perturbation $\epsilon$ along the gradient direction will influence the loss more if there is a larger gradient magnitude. 
In the white-box attack scenario, the attacker can easily attack the classifiers with the largest gradient magnitude on the original input. 
We use $g$ to represent the mean value of the gradient magnitudes of all members, 
and use $\nabla J_i$ as the gradient of the $i^{th}$ member in the ensemble. 
According to the negative correlation principle, we make the gradient magnitude of the $i^{th}$ member and the ensemble gradient magnitude 
negatively correlated with the gradient magnitude of other members in the ensemble.
After the training, the gradient magnitude of all members in the ensemble will be negatively correlated. 
This means that when the gradient magnitude of one member in the ensemble decrease, the gradient magnitude of other members will increase or decrease. 
However, no matter how the gradient magnitude of a single member changes, 
after all members in the ensemble conduct simultaneous interactive training through negative correlation, 
each member will get the best gradient magnitude over dataset.

Therefore, we propose the second regularization term, which is shown in Formula \eqref{negative correlation gradients}.

 
\begin{equation}\label{negative correlation gradients}
	\operatorname{Loss}_{\mathrm{norm}}=\frac{1}{g^2} \left(\left\|\nabla J_i\right\|_{2}- g\right) \sum_{j \neq i}^k \left(\left\|\nabla J_j\right\|_{2}-g\right)
\end{equation}

\begin{equation}
	g=\frac{1}{k} \sum_{i=1}^{\text{k}}\|\nabla J_i\|_{2}
\end{equation}

Where $g$ is the mean value of the gradient magnitudes of all members, $k$ is the number of members in the ensemble and $\nabla J_i$ represents the gradient of the $i^{th}$ member. 
We use the $L_2$ norm of the gradient to calculate the gradient magnitude of each member in the ensemble. 
By minimizing the regularization term, the gradient magnitudes of all members in the ensemble is negatively correlated, 
which means different members can learn better over different features.

\begin{figure}[h]
	\centering
	\includegraphics[width=.4\textwidth]{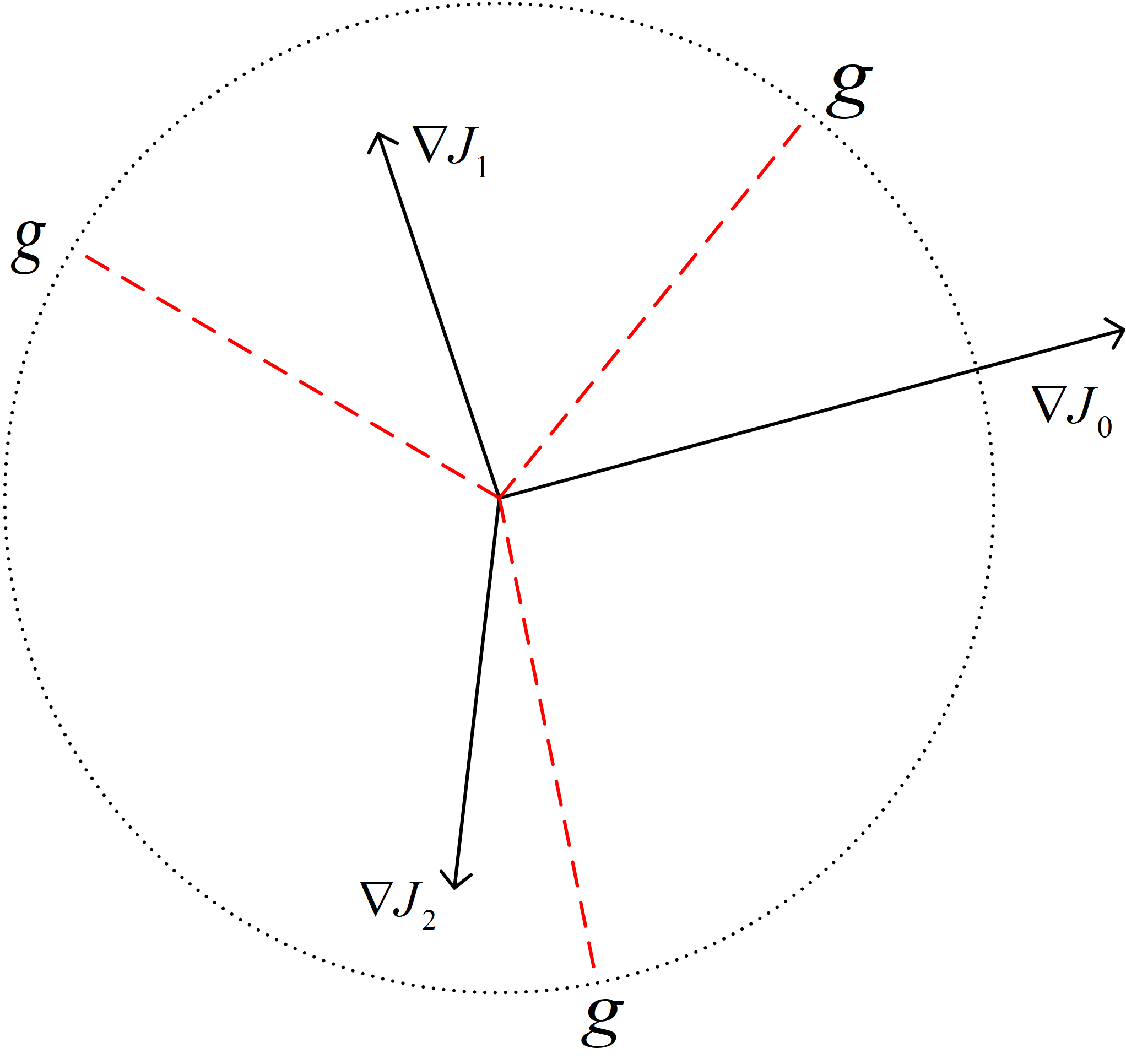} 
	\caption{Improving the ensemble diversity from the view of the gradient magnitude; $\nabla J_i$ indicates model gradients in the ensemble ($i=0,1,2$), and $g$ indicates the mean value of the gradient magnitudes of all members in the ensemble} 
	\label{length} 
\end{figure}

As shown in Fig. \ref{length}, where $g$ is the mean value of the gradient magnitudes of all members in the ensemble, 
and there must be some gradient magnitudes less than $g$, so it can be inferred that the value of the regularization term will always be negative. 
Based on this, we can further observe that:
\begin{itemize}
	\item When $\sum_{j \neq i}^k \left(\left\|\nabla J_j\right\|_{2}-g\right)$ is greater than zero, $\left\|\nabla J_i\right\|_{2}$ is less than $g$. 
	In such a situation, minimizing the negative correlation formula (\ref{negative correlation gradients}) will result in the decease of the gradient magnitude $\left\|\nabla J_i\right\|_{2}$.
	\item When $\sum_{j \neq i}^k \left(\left\|\nabla J_j\right\|_{2}-g\right)$ is negative, $\left\|\nabla J_i\right\|_{2}$ is greater than $g$. 
	In such a situation, minimizing the negative correlation formula (\ref{negative correlation gradients}) will result in the increase of the gradient magnitude $\left\|\nabla J_i\right\|_{2}$.
\end{itemize}

Considering all members are trained simultaneously, different members can learn different features over the training dataset and have 
different gradient magnitudes on the same input since it is not easy to always own very small gradient magnitudes over the whole input space. 
Based on this, it can be avoided that the adversaries attack the ensemble successfully by attacking only a few models with larger gradient magnitude. 

We use this to regularize the training for improving the ensemble diversity and the loss function can be expressed as Formula \eqref{loss norm}. 
Where $CE$ is the cross-entropy loss and $\lambda_{norm}$ is the weight coefficient.

\begin{equation}\label{loss norm}
	\text {Loss}=CE+\lambda_{\text {norm}} \text {Loss}_{\text {norm}}
\end{equation}

The algorithm for calculating the $Loss_{norm}$ with negative correlation is shown in Algorithm \ref{alg: norm},
where $k$ is the number of members in ensemble. 
Firstly, we calculate the difference $norm1$ between the gradient magnitude of the $i^{th}$ member and the mean value $g$ 
as well as the difference $norm2$ between the gradient magnitude of other members in the ensemble and the mean value $g$. 
After this, according to the negative correlation method, the $Loss_{norm}$ of $norm1$ and $norm2$ is calculated and treated as a regularization term for the training.

\begin{algorithm}[!ht]
	\caption{Calculating $Loss_{norm}$ for the $i^{th}$ member}
	\label{alg: norm}
	\begin{algorithmic}
		\Require
		Gradient of the $i^{th}$ member: $\nabla J_i$;
		Gradient of other members in the ensemble: $\nabla J_j(j \neq i)$;
		the mean value of the gradient magnitude: $g$
		\Ensure
		Regularization term: $Loss_{norm}$
		
		\State $norm1 \gets \left(\|\nabla J_i\|_2 - g\right)/g$
		\State $norm2 \gets 0$
		\For{$j \gets 1 \; \textbf{to} \; k$}
			\If {$j \neq i$}
				\State $norm2 \gets norm2 + (\|\nabla J_j\|_2 - g)/g$
			\EndIf
		\EndFor
		\State $Loss_{norm} \gets norm1 * norm2$
		\State Return the regularization term: $Loss_{norm}$
	\end{algorithmic}
\end{algorithm}

\subsection{The Proposed NCEn}
\label{sec:Joint diversity of ensemble and gradient changes of ensemble}
We have proposed two methods to increase the adversarial robustness, however, we find that using each method alone is still limites in performance. 

Specifically, if only considering the influence of the gradient direction of each member in the ensemble, 
an adversary can attack a few members with large gradient magnitudes to make the ensemble predict incorrectly.   
While if only considering the influence of the gradient magnitude of each member in the ensemble, 
there could be a phenomenon that the gradient directions of all members are similar.
At this time, the loss of each member in the ensemble will grow positively corrected,
i.e, adversarial examples generated along the gradient direction can make most members of the ensemble predict incorrectly. 
So that the defense performance of the ensemble is similar to that of a single model, which is not robust enough.

Therefore, we will consider both the influence of member gradient directions and the influence of member gradient magnitudes, 
and use two regularization terms simultaneously to improve the adversarial robustness.
The loss function is shown in Formula \eqref{loss}, 
where $CE$ is cross-entropy loss and the combined regularization term $NCE$ can be calculated by Formula \eqref{eq: nce}.

\begin{equation}\label{loss}
	\text{Loss}=CE+NCE
\end{equation}
\begin{equation}\label{eq: nce}
	NCE=\frac{1}{k} \sum_{i=1}^{k} \left( \lambda_{\text{norm}} \text{Loss}_{\text{norm}} + \lambda_{\text{cos}} \text{Loss}_{\text{cos}}\right)
\end{equation}

The specific implementation of the ensemble training process is shown in Algorithm \ref{alg: NCE}.
In each epoch, we first get the predicted value $pred$ of each member in the ensemble.
Next, we use the predicted value $pred$ and the real label $label$ to calculate the cross entropy loss $ce_{loss}$ of each member in the ensemble.
Then, we calculate the mean value $CE$ of the cross-entropy loss of all members in the ensemble. 
Finally, we use Formula \eqref{eq: nce} to calculate the regularization term $NCE$,
and by which we can update the parameters of all members.

\begin{algorithm}[!ht]
	\caption{Negative Correlation Ensemble}
	\label{alg: NCE}
	\begin{algorithmic}
		\Require
		Dataset: $X$;
		Correct label for dataset $X$: $label$
		\Ensure
		Trained ensemble

		\State Get a list of all the members in ensemble: $f_i (i=1,2,...,k)$
	
		\For{$epoch \gets begin\_epoch$ \textbf{to} $end\_epoch$}
			\For{$model \gets f_1$ \textbf{to} $f_k$}
				\State $pred \gets model(X)$
				\State $ce_{loss} \gets CE(pred, label)$
			\EndFor
			\State Calculate $CE \gets \frac{1}{k} \sum_{i=1}^{k}ce_{loss}$
			\State Get the value of the regularization term: $NCE$
			\State Back propagation using the sum of $CE$ and $NCE$
			\State Update the model parameters of each model
		\EndFor
	\end{algorithmic}
\end{algorithm}

\section{Experiments}
\label{sec:Experiment}
In this section, we conduct experiments on the FashionMNIST and CIFAR-10 datasets. 
We first give our experimental settings. 
Secondly, we compare the adversarial robustness of NCEn with different ensemble defense approaches. 
Furthermore, we find that NCEn is capable to reduce the transferability of adversarial examples, 
which proves that NCEn leverages the interactions of different members to improve the adversarial robustness.
The source codes of this work are available at \url{https://github.com/MiLabHITSZ/2022ZhangNCEn}

\subsection{Experimental Setup}
\label{setup}
In our experiments, we use three ensemble defense strategies as baselines to evaluate the performance of NCEn. 
The first one is the ensemble training without any regularization term, which $\lambda_{cos} = \lambda_{norm} = 0$. 
The second one is to use GAL for diversified training to improve the adversarial robustness of the ensemble \cite{2019Improving}.
The third one is the GPMR proposed by Dabouei et al. \cite{2020Exploiting}, 
which constructs the first-order defense interaction of the members in the ensemble to improve the adversarial robustness of the ensemble. 
Both GAL and GPMR are state-of-the-art ensemble defense strategies.

In our experiments, we use $k=3,4,5$ models in the ensemble respectively for analyzing the impact of the number of members, as shown in Table \ref{tab1}.
We evaluate the adversarial robustness of the ensemble on the FashionMNIST and CIFAR-10 datasets.
For GAL, the coefficient of the regularization term is set to 0.5. 
For GPMR, $\lambda_{div}$ of FashionMNIST is set to 0.1, $\lambda_{div}$ of CIFAR-10 is set to 0.04, and $\lambda_{eq}$ is set to 10 for all datasets. 

\begin{table}[!t]
	\centering
	\caption{Ensemble model structures. $k=(3,4,5)$}
	\small
	\label{tab1}
	\begin{tabular}{cm{6cm}}
		\toprule
		Name& Ensemble Structures\\
		\midrule
			Ensemble1& $k*ResNet20$\\
			Ensemble2& $k*ResNet26$\\
			Ensemble3& $k*ResNet32$\\
			Ensemble4& $(k-\lfloor \frac{k-1}{2} \rfloor)*ResNet20 + \lfloor \frac{k-1}{2} \rfloor *ResNet26 + 1*ResNet32$\\
		\bottomrule
	\end{tabular}
\end{table}

We train the ensemble with Adam optimizer \cite{kingma2014adam}.
The initial learning rate is set to $10^{-3}$, and decays with the factor of 0.1 every 15 epochs until reaching the final learning rate $10^{-5}$.
We train 40 epochs on FashionMNIST, and 60 epochs on CIFAR-10.
In the experiments, the batch size of all training processes is set to 64.
We set $\lambda_{cos} = 0.02$ and $\lambda_{norm} = 0.02$ for FashionMNIST,
$\lambda_{cos} = 0.06$ and $\lambda_{norm} = 0.04$ for CIFAR-10.

In FashionMNIST, we use random cropping, and in CIFAR-10 we use random horizontal flipping and random cropping. 
In addition, we add a new dataset $D_{noise}$ to the dataset D, which is generated by the perturbation from the truncated normal distribution: 
$N(\mu=0,\sigma=\epsilon/2)$, $ \epsilon=0.3$ for FashionMNIST and $\epsilon= 0.09$ for CIFAR-10. 
We use the synthetic data set $D +D_{noise}$ to train the ensemble. 
Such an operation can effectively distort the useless high-frequency features and prompt the ensemble to avoid overfitting during the training.

We use test accuracy of clean examples (ACE) and test accuracy of adversarial examples (AAE) as the evaluation metrics. 
AAE refers to the ratio that a new example set, which is constructed by adding adversarial perturbations to the clean examples, 
can still be correctly classified by the ensemble.
We use the averaged $CE$ loss of all models in the ensemble as the objective function of the attack. 

\subsection{Defense Performance}
\label{White box defense performance}
We compare the defense performance of NCEn with different defense methods, as shown in Tables \ref{num3}$\sim$\ref{num5}, 
and the best defense results are bolded. 
We use several powerful white box attack methods to test the defense performance of ensembles, 
including the fast gradient sign method (FGSM) \cite{Goodfellow2015}, the momentum iterative fast gradient sign method (MI-FGSM) \cite{dong2018boosting}, 
projected gradient descent (PGD) \cite{DBLP:conf/iclr/MadryMSTV18}, the basic iterative method (BIM) \cite{Kurakin2019}.
The specific settings of attack methods are as follows.
FGSM adds a perturbation with a step length of $\epsilon$ in the gradient direction. 
For MI-FGSM, each step size is set to 0.01, while keeping the maximum distortion always within $\epsilon$ of the initial point. 
For PGD, the number of iterations is set to $40$, and the maximum perturbation and single-step attack steps are both $\epsilon$. 
For BIM, the number of iterations is set to $10$, and the maximum perturbation and single-step attack steps are both $\epsilon$. 
All attack methods are implemented by AdveTorch.

When the number of models in the ensemble is $3$, 
the results of different approaches on dataset FashionMNIST$/$CIFAR-10 are shown in Table \ref{num3}. 
Compared with BL, GAL and GPMR, the adversarial robustness of NCEn is significantly improved. 
When we change the ensemble model structure to Ensemble2, Ensemble3 and Ensemble4 respectively, the ACE of GAL drops below $0.8$. 
At this time, the AAE of GAL is unworthy of consideration. 
Such AAE are marked with gray background as well as the corresponding ACE
Unlike GAL, NCEn can not only maintain high ACE in all ensemble model structures, but also achieve better adversarial robustness. 
Specifically, on FashionMNIST, except that FGSM and MI-FGSM use a large attack step, NCEn has better adversarial robustness in all ensemble structures. 
AAE has the highest improvement of $0.14$ on Ensemble4 and MI-FGSM $(\epsilon=0.1)$. 
On CIFAR-10, the adversarial robustness of NCEn is better or very close to the existing best baselines, and AAE can be improved by $0.11$ at most.

When the number of models in the ensemble is $4$ or $5$, the results of different defense strategies on FashionMNIST$/$CIFAR-10 are shown in Tables \ref{num4}$\sim$\ref{num5}. 
The experimental results are consistent with the previous ones, and NCEn still has better adversarial robustness. 
Specifically, when the number of models in the ensemble is $4$, the AAE score of NCEn can be increased by $0.12$ on FashionMNIST. 
On CIFAR-10, the AAE of NCEn is higher than all baselines, and the highest AAE can be improved by $0.12$. 
When the number of models in the ensemble is $5$, the AAE of NCEn can be increased by $0.12$ on FashionMNIST. 
On CIFAR-10, the highest AAE of NCEn can be improved by $0.11$.

\subsection{Transferability Between Members}
\label{Transferability between members}

\begin{figure*}[htbp]
	\centering
	\subfigure{
	\includegraphics[width=.23\textwidth]{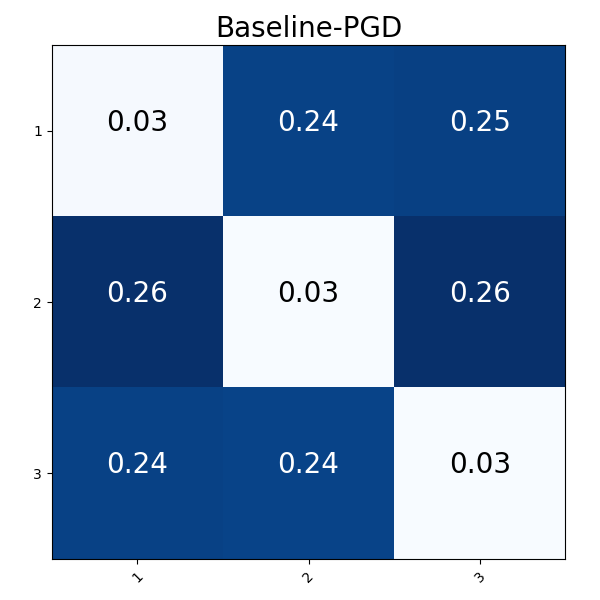}
	}
	\subfigure{
	\includegraphics[width=.23\textwidth]{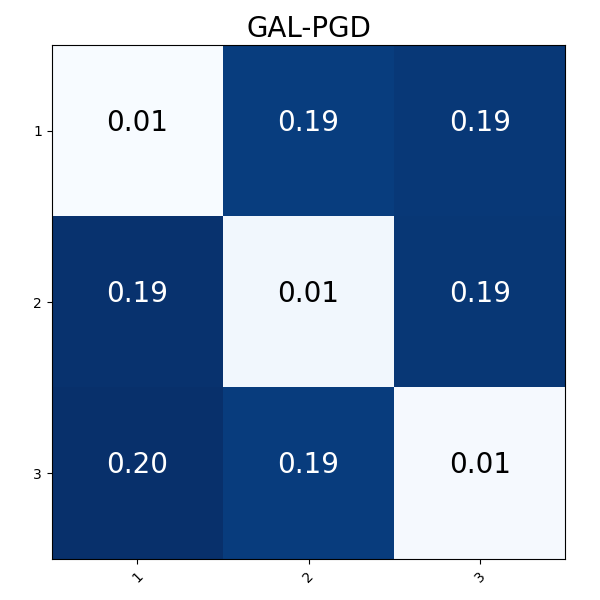}
	}
	\subfigure{
	\includegraphics[width=.23\textwidth]{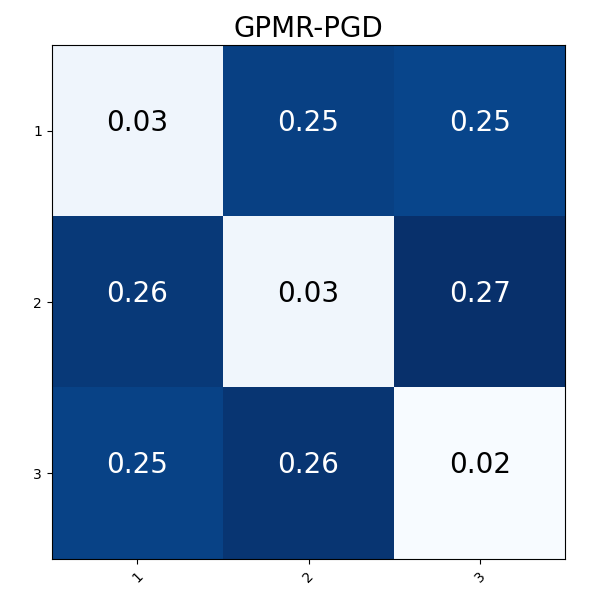}
	}
	\subfigure{
	\includegraphics[width=.23\textwidth]{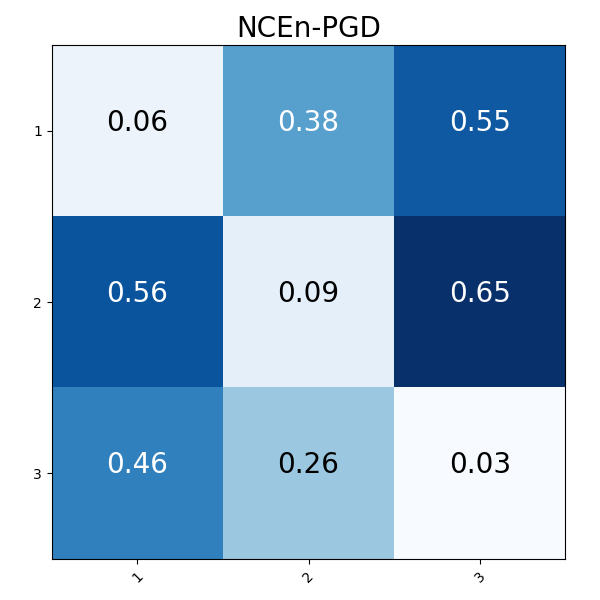}
	}

	\subfigure{
	\includegraphics[width=.23\textwidth]{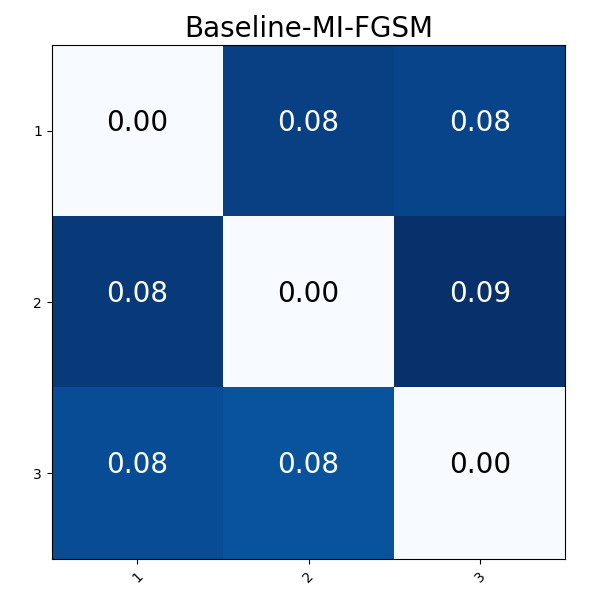}
	}
	\subfigure{
	\includegraphics[width=.23\textwidth]{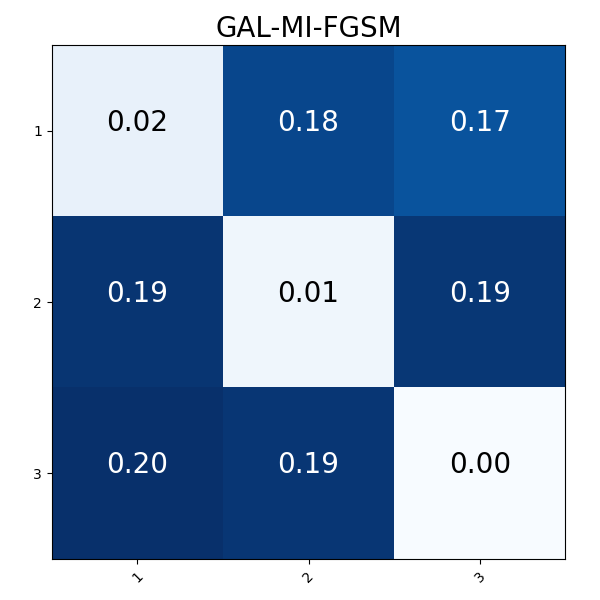}
	}
	\subfigure{
	\includegraphics[width=.23\textwidth]{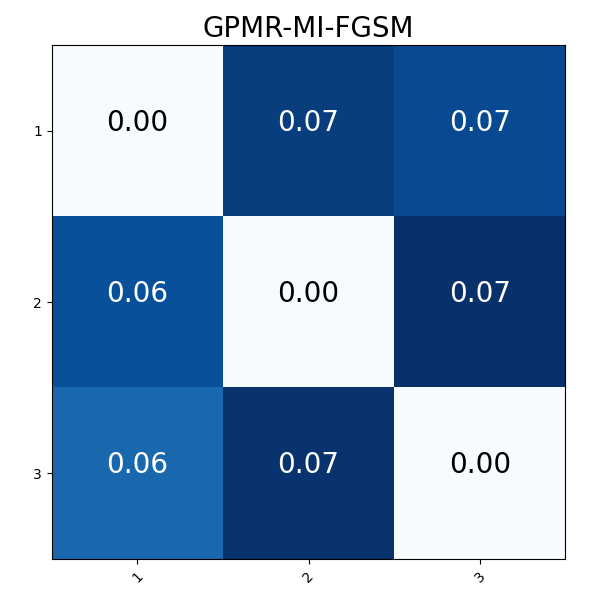}
	}
	\subfigure{
	\includegraphics[width=.23\textwidth]{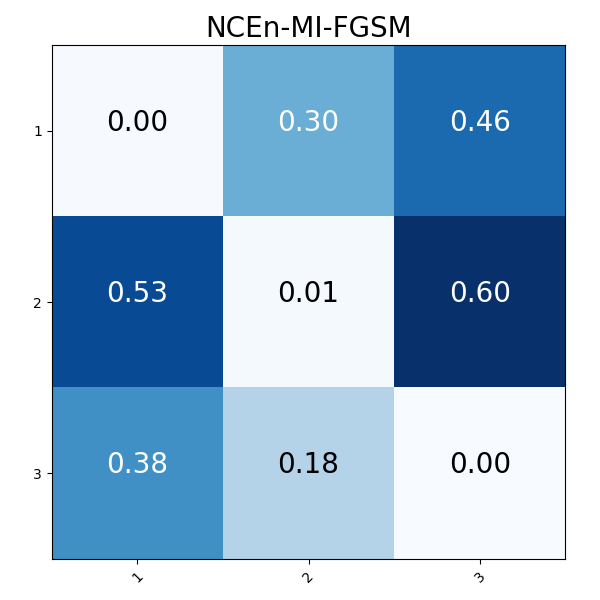}
	}
	\caption{The confusion matrix of the accuracy of the adversarial examples test on CIFAR-10 by each member in the ensemble. 
	Rows and columns respectively represent the model for generating adversarial examples and the model for testing the accuracy of adversarial examples.}
	\label{Transferability}
\end{figure*} 

Transferability refers to the success rate of being able to attack other models at the same time when an adversarial example is designed to attack a 
particular model. 
The less transferable the adversarial example is, the better the diversity of the models

Therefore, we can use the transferability of adversarial examples between different models to evaluate the similarity between members in the ensemble.
We perform transferability experiments using PGD \cite{DBLP:conf/iclr/MadryMSTV18} and MI-FGSM \cite{dong2018boosting},
which MI-FGSM performs well in black-box attacks and PGD is the most powerful first-order attack algorithm. 
We generate adversarial examples for each member, and then evaluate their transferability on other members by calculating AAE.
The perturbation magnitude of all attacks is set to $\epsilon=0.05$, we use Ensemble4 in Table \ref{tab1} to conduct the transferability experiments,
and the results are shown in Fig. \ref{Transferability}.

As shown in Fig. \ref{Transferability}, we use the heat map of the confusion matrix to show transferability.
The $i$th row and $j$th column in the heat map represents the test accuracy of the adversarial examples on the $j$th member, while the adversarial examples are generated over the $i$th member.
When the values of other positions in the confusion matrix are close to those of the diagonal, 
it means that the adversarial examples generated by the $i$th model can successfully attack other models in the ensemble.
The closer the values in the confusion matrix are, the higher the transferability of the adversarial examples and the higher the similarity between members in the ensemble.
On the contrary, it means that the diversity among members in the ensemble is higher, and the ensemble has better adversarial robustness.
It can be seen from the confusion matrix that the transferability of adversarial examples in NCEn ensemble is poor, 
which indicates that NCEn has better diversity, such that it can provide better defense interactions for members.

\subsection{Parameter Analysis}
\label{parameter analysis}

\begin{figure}[h]
	\includegraphics[width=.4\textwidth]{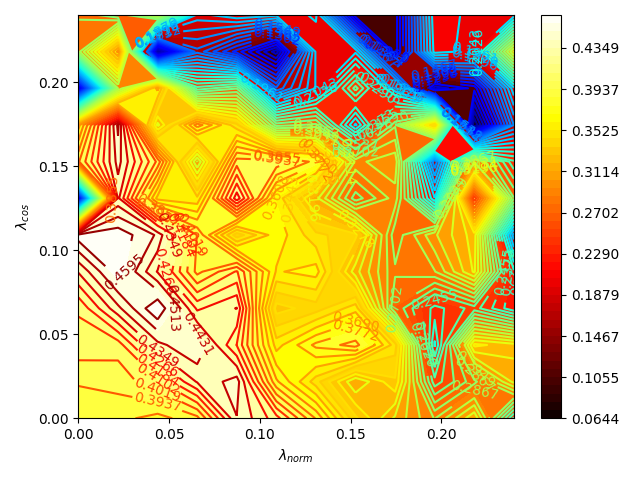} 
	\caption{Contour Heat Map of $\lambda_{cos}$ and $\lambda_{norm}$}
	\label{hdd} 
\end{figure}

From the experiment, we can find that the weights of the  regularization terms will affect the defense performance. 
In order to explore the relationship between the weights and the defense performance, 
we summarize the experimental results and plot the contour heat map on $\lambda_{cos}$ and $\lambda_{norm}$ (see Figure \ref{hdd}).
We use the product of ACE and AAE (which is averaged over four attack methods) in Table \ref{num3} to represent the intensities of the heat map. 
The larger the intensity, the higher the attack success rate.
We use Ensemble1 in Table \cref{tab1} and test on CIFAR-10.
From Figure \ref{hdd}, we can get that the best results are obtained when we set $\lambda_{cos}=0.06$ and $\lambda_{norm}=0.04$.

\begin{table*}[!th]
	\centering
	\caption{the classification accuracy of four ensemble defense strategies adversarial examples on FashionMNIST and CIFAR-10, when model number is 3. ACE means accuracy of clean examples.}
	\label{num3}
	\resizebox{.9\textwidth}{!}{
	\begin{tabular}{c|c|c|cccc|c|cccc}
		\toprule
		\multirow{2}{*}{Ensemble}	&\multirow{2}{*}{Attack}	& \multirow{2}{*}{Setting}	& \multicolumn{4}{c|}{FashionMNIST}& \multirow{2}{*}{Setting}	&\multicolumn{4}{c}{CIFAR-10}\\
		&&&BL&GAL&GPMR&NCEn&&BL&GAL&GPMR&NCEn\\
		\cline{1-12}
		\multirow{8}{*}{Ensemble1}
		&&\multicolumn{1}{c|}{ACE}	& 0.9337	& 0.8925& 0.9205& 0.9184	&\multicolumn{1}{c|}{ACE}&0.8639& 0.8070& 0.8433& 0.8436\\
		\cline{2-12}
		&\multirow{2}{*}{FGSM}		&$\epsilon$=0.1		& 0.8459& 0.8168& 0.8178& \textbf{0.8656} 					&$\epsilon$=0.03& 0.6744& 0.6102& 0.6118& \textbf{0.7473}\\
									&&$\epsilon$=0.3	& 0.6131& \textbf{0.6885}& 0.6847& 0.6414 					&$\epsilon$=0.09& 0.5462& 0.4421& 0.4986& \textbf{0.6217}\\
									\cline{2-12}
		&\multirow{2}{*}{MI-FGSM}	&$\epsilon$=0.1		& 0.6310& 0.5253& 0.5720& \textbf{0.6882}					&$\epsilon$=0.03& 0.2067& 0.0131& 0.1530& \textbf{0.3000}\\
									&&$\epsilon$=0.3	& \textbf{0.4759}& 0.3396& 0.3867&0.4412 					&$\epsilon$=0.09& 0.0625& 0.0004& 0.0500& \textbf{0.1157}\\
									\cline{2-12}
		&\multirow{2}{*}{PGD}		&$\epsilon$=0.1		& 0.7002& 0.6406& 0.5982& \textbf{0.7694}					&$\epsilon$=0.01& 0.5661& 0.5708& 0.4963& \textbf{0.6473}\\	
									&&$\epsilon$=0.15	& 0.6163& 0.5434& 0.4983& \textbf{0.7362}					&$\epsilon$=0.02& 0.4307& 0.2117& 0.3459& \textbf{0.5305}\\
									\cline{2-12}
		&\multirow{2}{*}{BIM}		&$\epsilon$=0.1		& 0.7079& 0.6674& 0.6291& \textbf{0.7783}					&$\epsilon$=0.01& 0.6119& 0.5382& 0.5241& \textbf{0.6740}\\
									&&$\epsilon$=0.15	& 0.6292& 0.5620& 0.5264& \textbf{0.7383}					&$\epsilon$=0.2& 0.4722& 0.2469& 0.3916& \textbf{0.5716}\\
									\cline{1-12}
		&&\multicolumn{1}{c|}{ACE}	&	0.9326& 	\cellcolor{gray}0.6724& 	0.9203& 	0.9208	&\multicolumn{1}{c|}{ACE}&0.8688& \cellcolor{gray}0.7842& 0.8385& 0.8571\\
		\cline{2-12}
		\multirow{8}{*}{Ensemble2}
		&\multirow{2}{*}{FGSM}		&$\epsilon$=0.1		& 0.8396& 	\cellcolor{gray}0.9284& 	0.7995& 	\textbf{0.8525} 			&$\epsilon$=0.03& \textbf{0.6862}& \cellcolor{gray}0.6884& 0.5933& 0.6844\\
									&&$\epsilon$=0.3	& 0.6025& 	\cellcolor{gray}0.7754& 	\textbf{0.6975}& 	0.6564 				&$\epsilon$=0.09& \textbf{0.5606}& \cellcolor{gray}0.5673& 0.4697& 0.5570\\
									\cline{2-12}
		&\multirow{2}{*}{MI-FGSM}	&$\epsilon$=0.1		& 0.6155& 	\cellcolor{gray}0.7751& 	0.5605&		\textbf{0.6725}				&$\epsilon$=0.03& 0.2107&\cellcolor{gray} 0.0175& 0.1404& \textbf{0.2405}\\
									&&$\epsilon$=0.3	& 0.4461& 	\cellcolor{gray}0.4531& 	0.4018& 	\textbf{0.4549}				&$\epsilon$=0.09& 0.0709&\cellcolor{gray} 0.0051& 0.0374& \textbf{0.0851}\\
									\cline{2-12}
		&\multirow{2}{*}{PGD}		&$\epsilon$=0.1		& 0.6904& 	\cellcolor{gray}0.8334& 	0.6014& 	\textbf{0.7516}				&$\epsilon$=0.01& 0.5719&\cellcolor{gray} 0.5728& 0.4819& \textbf{0.6013}\\	
									&&$\epsilon$=0.15	& 0.6210& 	\cellcolor{gray}0.7360&		0.4883&		\textbf{0.6917}				&$\epsilon$=0.02& 0.4391&\cellcolor{gray} 0.3555& 0.3378& \textbf{0.4735}\\
									\cline{2-12}
		&\multirow{2}{*}{BIM}		&$\epsilon$=0.1		& 0.7060& 	\cellcolor{gray}0.8511& 	0.6220& \textbf{0.7671}					&$\epsilon$=0.01& 0.6109&\cellcolor{gray} 0.5618& 0.5197& \textbf{0.6472}\\
									&&$\epsilon$=0.15	& 0.6107& 	\cellcolor{gray}0.7473& 	0.5182&		\textbf{0.6933}				&$\epsilon$=0.2& 0.4762&\cellcolor{gray} 0.3816& 0.3831& \textbf{0.5061}\\
									\cline{1-12}
		&&\multicolumn{1}{c|}{ACE}	&	0.9316&\cellcolor{gray}	0.5259&	0.9198&	0.9212	&\multicolumn{1}{c|}{ACE}&0.8665&\cellcolor{gray} 0.7589& 0.8387& 0.8408\\
		\cline{2-12}
		\multirow{8}{*}{Ensemble3}
		&\multirow{2}{*}{FGSM}		&$\epsilon$=0.1		& 0.8535&\cellcolor{gray}	0.8836&		0.8259&		\textbf{0.8782} 				&$\epsilon$=0.03& 0.6873&\cellcolor{gray} 0.5577& 0.6076& \textbf{0.7614}\\
									&&$\epsilon$=0.3	& 0.6327&\cellcolor{gray}	0.7649&	\textbf{0.6797}&	0.6651 						&$\epsilon$=0.09& 0.5644&\cellcolor{gray} 0.4507& 0.4864& \textbf{0.6321}\\
									\cline{2-12}
		&\multirow{2}{*}{MI-FGSM}	&$\epsilon$=0.1		& 0.6456&\cellcolor{gray}	0.6280&	0.5842&	\textbf{0.7028}							&$\epsilon$=0.03& 0.2148&\cellcolor{gray} 0.0052& 0.1423& \textbf{0.3159}\\
									&&$\epsilon$=0.3	& 0.4913&\cellcolor{gray}	0.4772&	0.4016&	\textbf{0.5593}							&$\epsilon$=0.09& 0.0679&\cellcolor{gray} 0.0006& 0.0337& \textbf{0.1202}\\
									\cline{2-12}
		&\multirow{2}{*}{PGD}		&$\epsilon$=0.1		& 0.7075&\cellcolor{gray}	0.7837&	0.6133&	\textbf{0.7871}							&$\epsilon$=0.01& 0.5764&\cellcolor{gray} 0.3464& 0.4914& \textbf{0.6678}\\	
									&&$\epsilon$=0.15	& 0.6412&\cellcolor{gray}	0.6773&	0.5310&	\textbf{0.7147}							&$\epsilon$=0.02& 0.4482&\cellcolor{gray} 0.2582& 0.3369& \textbf{0.5448}\\
									\cline{2-12}
		&\multirow{2}{*}{BIM}		&$\epsilon$=0.1		& 0.7183&\cellcolor{gray}	0.7961&	0.6428&	\textbf{0.7864}							&$\epsilon$=0.01& 0.6168&\cellcolor{gray} 0.3390& 0.5218& \textbf{0.6912}\\
									&&$\epsilon$=0.15	& 0.6384&\cellcolor{gray}	0.7003&	0.5472&	\textbf{0.7206}							&$\epsilon$=0.2& 0.4859&\cellcolor{gray} 0.2205& 0.3851& \textbf{0.5936}\\
									\cline{1-12}
		&&\multicolumn{1}{c|}{ACE}	&	0.9309&\cellcolor{gray} 	0.7503& 	0.9205& 	0.8428	&\multicolumn{1}{c|}{ACE}&0.8695&\cellcolor{gray}	0.6851&	0.8426&	0.8492\\
		\cline{2-12}
		\multirow{8}{*}{Ensemble4}
		&\multirow{2}{*}{FGSM}		&$\epsilon$=0.1		& 0.8505&\cellcolor{gray} 	0.9338& 	0.8116& 	\textbf{0.8992} 				&$\epsilon$=0.03& 0.6813&\cellcolor{gray}	0.6293&		0.6011&		\textbf{0.7081}\\
									&&$\epsilon$=0.3	& 0.6319&\cellcolor{gray} 	0.8501& 	0.6793& 	\textbf{0.7257} 				&$\epsilon$=0.09& 0.5587&\cellcolor{gray}	0.4936&		0.4831&		\textbf{0.5827}\\
									\cline{2-12}
		&\multirow{2}{*}{MI-FGSM}	&$\epsilon$=0.1		& 0.6360&\cellcolor{gray} 	0.7722& 0.6015&	\textbf{0.7734}							&$\epsilon$=0.03& 0.2118&\cellcolor{gray}		0.1224&		0.1482&		\textbf{0.2731}\\
									&&$\epsilon$=0.3	& 0.4820&\cellcolor{gray} 	0.5316& 	0.4285& 	\textbf{0.5202}					&$\epsilon$=0.09& 0.0672&\cellcolor{gray}		0.0389&		0.0431&		\textbf{0.0959}\\
									\cline{2-12}
		&\multirow{2}{*}{PGD}		&$\epsilon$=0.1		& 0.6954&\cellcolor{gray} 	0.8527& 	0.6160& 	\textbf{0.8054}					&$\epsilon$=0.01& 0.5738&\cellcolor{gray}		0.5831&		0.4897&		\textbf{0.6295}\\	
									&&$\epsilon$=0.15	& 0.6235&\cellcolor{gray} 	0.7651& 	0.5177& 	\textbf{0.7469}					&$\epsilon$=0.02& 0.4426&\cellcolor{gray}		0.3595&		0.3332&		\textbf{0.4981}\\
									\cline{2-12}
		&\multirow{2}{*}{BIM}		&$\epsilon$=0.1		& 0.7073&\cellcolor{gray} 	0.8571& 	0.6337& 	\textbf{0.8156}					&$\epsilon$=0.01& 0.6165&\cellcolor{gray}	0.5739&		0.5185&		\textbf{0.6637}\\
									&&$\epsilon$=0.15	& 0.6231&\cellcolor{gray} 	0.7654& 	0.5434& 	\textbf{0.7543}					&$\epsilon$=0.2& 0.4833&\cellcolor{gray}		0.3908&		0.3876&		\textbf{0.5449}\\
		\bottomrule
	\end{tabular}}
\end{table*}

\begin{table*}[!th]
	\centering
	\caption{the classification accuracy of four ensemble defense strategies adversarial examples on FashionMNIST and CIFAR-10, when model number is 4. ACE means accuracy of clean examples.}
	\label{num4}
	\resizebox{.9\textwidth}{!}{
	\begin{tabular}{c|c|c|cccc|c|cccc}
		\toprule
		\multirow{2}{*}{Ensemble}	&\multirow{2}{*}{Attack}	& \multirow{2}{*}{Setting}	& \multicolumn{4}{c|}{FashionMNIST}& \multirow{2}{*}{Setting}	&\multicolumn{4}{c}{CIFAR-10}\\
		&&&BL&GAL&GPMR&NCEn&&BL&GAL&GPMR&NCEn\\
		\cline{1-12}
		\multirow{8}{*}{Ensemble1}
		&&\multicolumn{1}{c|}{ACE}	&	0.9340& 0.9049& 0.9235& 0.9209	&\multicolumn{1}{c|}{ACE}&0.8659& 0.8055& 0.8527& 0.8494\\
		\cline{2-12}
		&\multirow{2}{*}{FGSM}		&$\epsilon$=0.1		& 0.8455& 0.8636& 0.8351& \textbf{0.8708} 					&$\epsilon$=0.03& 0.6642& 0.6788& 0.6010& \textbf{0.7356}\\
									&&$\epsilon$=0.3	& 0.6267& \textbf{0.7506}& 0.7090&0.6523  					&$\epsilon$=0.09& 0.5445& 0.5333& 0.4768& \textbf{0.6074}\\
									\cline{2-12}
		&\multirow{2}{*}{MI-FGSM}	&$\epsilon$=0.1		& 0.6358& 0.6118& 0.6021& \textbf{0.6940}					&$\epsilon$=0.03& 0.1984& 0.0977& 0.1470& \textbf{0.2727}\\
									&&$\epsilon$=0.3	& \textbf{0.4867}& 0.4050& 0.4442& 0.4364					&$\epsilon$=0.09& 0.0605& 0.0022& 0.0418& \textbf{0.0996}\\
									\cline{2-12}
		&\multirow{2}{*}{PGD}		&$\epsilon$=0.1		& 0.6784& 0.6670& 0.5916& \textbf{0.7620}					&$\epsilon$=0.01& 0.5505& 0.6216& 0.5013& \textbf{0.6266}\\	
									&&$\epsilon$=0.15	& 0.5962& 0.5414& 0.5046& \textbf{0.6968}					&$\epsilon$=0.02& 0.4072& 0.2813& 0.3342& \textbf{0.4901}\\
									\cline{2-12}
		&\multirow{2}{*}{BIM}		&$\epsilon$=0.1		& 0.6928& 0.6894& 0.6321& \textbf{0.7690}					&$\epsilon$=0.01& 0.6003& 0.6300& 0.5323& \textbf{0.6612}\\
									&&$\epsilon$=0.15	& 0.6080& 0.5638& 0.5352& \textbf{0.7117}					&$\epsilon$=0.2& 0.4434& 0.3492& 0.3810& \textbf{0.5376}\\
									\cline{1-12}
		&&\multicolumn{1}{c|}{ACE}	&	0.9338& 	0.8913& 	0.9228& 	0.9043	&\multicolumn{1}{c|}{ACE}&0.8678&\cellcolor{gray} 0.7942& 0.8575& 0.8496\\
		\cline{2-12}
		\multirow{8}{*}{Ensemble2}
		&\multirow{2}{*}{FGSM}		&$\epsilon$=0.1		& 0.8421& 	\textbf{0.8521}& 	0.8333& 	0.8454 				&$\epsilon$=0.03& 0.6665&\cellcolor{gray} 0.7076& 0.6020& \textbf{0.7162}\\
									&&$\epsilon$=0.3	& 0.6081& 	\textbf{0.7198}& 	0.7190& 	0.6403 					&$\epsilon$=0.09& 0.5436&\cellcolor{gray} 0.5732& 0.4772& \textbf{0.5836}\\
									\cline{2-12}
		&\multirow{2}{*}{MI-FGSM}	&$\epsilon$=0.1		& 0.6243& 	0.5796& 	0.6111& 	\textbf{0.6806}					&$\epsilon$=0.03& 0.2014&\cellcolor{gray} 0.1126& 0.1498& \textbf{0.2704}\\
									&&$\epsilon$=0.3	& 0.4499& 	0.3944& 	\textbf{0.4642}& 	0.4490					&$\epsilon$=0.09& 0.0667&\cellcolor{gray} 0.0045& 0.0415& \textbf{0.0952}\\
									\cline{2-12}
		&\multirow{2}{*}{PGD}		&$\epsilon$=0.1		& 0.6835& 	0.6753& 	0.5988& 	\textbf{0.7421}					&$\epsilon$=0.01& 0.5706&\cellcolor{gray} 0.5849& 0.4997& \textbf{0.6394}\\	
									&&$\epsilon$=0.15	& 0.6019& 	0.5270&		0.5123&		\textbf{0.6836}					&$\epsilon$=0.02& 0.4136&\cellcolor{gray} 0.3283& 0.3369& \textbf{0.4871}\\
									\cline{2-12}
		&\multirow{2}{*}{BIM}		&$\epsilon$=0.1		& 0.6855& 	0.6977& 	0.6324& 	\textbf{0.7451}					&$\epsilon$=0.01& 0.6114&\cellcolor{gray} 0.5931& 0.5320& \textbf{0.6673}\\
									&&$\epsilon$=0.15	& 0.5999& 	0.5580& 	0.5409&		\textbf{0.6861}					&$\epsilon$=0.2& 0.4566&\cellcolor{gray} 0.3857& 0.3884& \textbf{0.5387}\\
									\cline{1-12}
		&&\multicolumn{1}{c|}{ACE}	&	0.9320&\cellcolor{gray}	0.7638&	0.9255&	0.8980	&\multicolumn{1}{c|}{ACE}&0.8718&\cellcolor{gray} 0.7911& 0.8537& 0.8504\\
		\cline{2-12}
		\multirow{8}{*}{Ensemble3}
		&\multirow{2}{*}{FGSM}		&$\epsilon$=0.1		& 0.8505&\cellcolor{gray}	0.8634&		0.8294&		\textbf{0.8678} 				&$\epsilon$=0.03& 0.6670&\cellcolor{gray} 0.7181& 0.6098& \textbf{0.7683}\\
									&&$\epsilon$=0.3	& 0.6493&\cellcolor{gray}	0.7052&	\textbf{0.6975}&	0.6347						&$\epsilon$=0.09& 0.5462&\cellcolor{gray} 0.6059& 0.4870& \textbf{0.6498} \\
									\cline{2-12}
		&\multirow{2}{*}{MI-FGSM}	&$\epsilon$=0.1		& 0.6433&\cellcolor{gray}	0.5943&	0.6128&	\textbf{0.7239}							&$\epsilon$=0.03& 0.2068&\cellcolor{gray} 0.1237& 0.1451& \textbf{0.3226}\\
									&&$\epsilon$=0.3	& 0.5037&\cellcolor{gray}	0.4409&	0.4534&	\textbf{0.5465}							&$\epsilon$=0.09& 0.0571&\cellcolor{gray} 0.0297& 0.0366& \textbf{0.1089}\\
									\cline{2-12}
		&\multirow{2}{*}{PGD}		&$\epsilon$=0.1		& 0.6979&\cellcolor{gray}	0.7526&	0.6024&	\textbf{0.7899}							&$\epsilon$=0.01& 0.5724&\cellcolor{gray} 0.5496& 0.4979& \textbf{0.6456}\\	
									&&$\epsilon$=0.15	& 0.6113&\cellcolor{gray}	0.6671&	0.5121&	\textbf{0.7097}							&$\epsilon$=0.02& 0.4183&\cellcolor{gray} 0.3541& 0.3409& \textbf{0.4995}\\
									\cline{2-12}
		&\multirow{2}{*}{BIM}		&$\epsilon$=0.1		& 0.7146&\cellcolor{gray}	0.7717&	0.6339&	\textbf{0.7884}							&$\epsilon$=0.01& 0.6097&\cellcolor{gray} 0.5729& 0.5314& \textbf{0.6880}\\
									&&$\epsilon$=0.15	& 0.6305&\cellcolor{gray}	0.6804&	0.5479&	\textbf{0.7119}							&$\epsilon$=0.2& 0.4652&\cellcolor{gray} 0.3953& 0.3863& \textbf{0.5569}\\
									\cline{1-12}
		&&\multicolumn{1}{c|}{ACE}	&	0.9327&\cellcolor{gray} 	0.7846& 	0.9224& 	0.9226	&\multicolumn{1}{c|}{ACE}&0.8706&\cellcolor{gray}	0.7961&	0.8532&	0.8668\\
		\cline{2-12}
		\multirow{8}{*}{Ensemble4}
		&\multirow{2}{*}{FGSM}		&$\epsilon$=0.1		& 0.8478&\cellcolor{gray} 	0.8817& 	0.8299& 	\textbf{0.8715}				&$\epsilon$=0.03& 0.6577&\cellcolor{gray}	0.7283&		0.6037&		\textbf{0.7530}\\
									&&$\epsilon$=0.3	& 0.6155&\cellcolor{gray} 	0.7931& 	\textbf{0.6823}& 	0.6622 				&$\epsilon$=0.09& 0.5342&\cellcolor{gray}	0.6036&		0.4731&		\textbf{0.6481}\\
									\cline{2-12}
		&\multirow{2}{*}{MI-FGSM}	&$\epsilon$=0.1		& 0.6264&\cellcolor{gray} 	0.6707& 	0.6013& 	\textbf{0.7145}					&$\epsilon$=0.03& 0.1952&\cellcolor{gray}		0.0771&		0.1468&		\textbf{0.2812}\\
									&&$\epsilon$=0.3	& \textbf{0.4726}&\cellcolor{gray} 	0.5034& 	0.4340& 	0.4640					&$\epsilon$=0.09& 0.0593&\cellcolor{gray}		0.0113&		0.0406&		\textbf{0.1016}\\
									\cline{2-12}
		&\multirow{2}{*}{PGD}		&$\epsilon$=0.1		& 0.6812&\cellcolor{gray} 	0.7178& 	0.5944& 	\textbf{0.7834}					&$\epsilon$=0.01& 0.5607&\cellcolor{gray}		0.5288&		0.5004&		\textbf{0.6435}\\	
									&&$\epsilon$=0.15	& 0.6043&\cellcolor{gray} 	0.6178& 	0.5079& 	\textbf{0.7241}					&$\epsilon$=0.02& 0.4121&\cellcolor{gray}		0.2905&		0.3363&		\textbf{0.5034}\\
									\cline{2-12}
		&\multirow{2}{*}{BIM}		&$\epsilon$=0.1		& 0.6959&\cellcolor{gray} 	0.7378& 	0.6201& 	\textbf{0.7818}					&$\epsilon$=0.01& 0.6046&\cellcolor{gray}	0.5415&		0.5303&		\textbf{0.6801}\\
									&&$\epsilon$=0.15	& 0.6096&\cellcolor{gray} 	0.6440& 	0.5388& 	\textbf{0.7194}					&$\epsilon$=0.2& 0.4501&\cellcolor{gray}		0.3220&		0.3809&		\textbf{0.5532}\\
		\bottomrule
	\end{tabular}}
\end{table*}

\begin{table*}[!th]
	\centering
	\caption{the classification accuracy of four ensemble defense strategies adversarial examples on FashionMNIST and CIFAR-10, when model number is 5. ACE means accuracy of clean examples. }
	\label{num5}
	\resizebox{.9\textwidth}{!}{
	\begin{tabular}{c|c|c|cccc|c|cccc}
		\toprule
		\multirow{2}{*}{Ensemble}	&\multirow{2}{*}{Attack}	& \multirow{2}{*}{Setting}	& \multicolumn{4}{c|}{FashionMNIST}& \multirow{2}{*}{Setting}	&\multicolumn{4}{c}{CIFAR-10}\\
		&&&BL&GAL&GPMR&NCEn&&BL&GAL&GPMR&NCEn\\
		\cline{1-12}
		\multirow{8}{*}{Ensemble1}
		&&\multicolumn{1}{c|}{ACE}	&	0.9315& 0.8998& 0.9247& 0.9235	&\multicolumn{1}{c|}{ACE}&0.8694& 0.8355& 0.8573& 0.8488\\
		\cline{2-12}
		&\multirow{2}{*}{FGSM}		&$\epsilon$=0.1		& 0.8446& 0.8546& 0.8330& \textbf{0.8619} 				&$\epsilon$=0.03& 0.6473& \textbf{0.7655}& 0.6100& 0.7248\\
									&&$\epsilon$=0.3	& 0.6265& \textbf{0.7482}& 0.7073& 0.6248					&$\epsilon$=0.09& 0.5258& \textbf{0.6266}& 0.4875& 0.6054\\
									\cline{2-12}
		&\multirow{2}{*}{MI-FGSM}	&$\epsilon$=0.1		& 0.6333& 0.6572& 0.6038& \textbf{0.6832}					&$\epsilon$=0.03& 0.1808& 0.0418& 0.1564& \textbf{0.2611}\\
									&&$\epsilon$=0.3	& \textbf{0.4813}& 0.3860& 0.4543& 0.4092					&$\epsilon$=0.09& 0.0512& 0.0021& 0.0422& \textbf{0.0919}\\
									\cline{2-12}
		&\multirow{2}{*}{PGD}		&$\epsilon$=0.1		& 0.6673& 0.6694& 0.6053& \textbf{0.7386}					&$\epsilon$=0.01& 0.5365& 0.5368& 0.4934& \textbf{0.6183}\\	
									&&$\epsilon$=0.15	& 0.5834& 0.5660& 0.5211& \textbf{0.6794}					&$\epsilon$=0.02& 0.3818& 0.2794& 0.3378& \textbf{0.4687}\\
									\cline{2-12}
		&\multirow{2}{*}{BIM}		&$\epsilon$=0.1		& 0.6809& 0.6984& 0.6350& \textbf{0.7524}					&$\epsilon$=0.01& 0.5845& 0.5770& 0.5273& \textbf{0.6528}\\
									&&$\epsilon$=0.15	& 0.5978& 0.5965& 0.5578& \textbf{0.6868}					&$\epsilon$=0.2& 0.4324& 0.3244& 0.3909& \textbf{0.5192}\\
									\cline{1-12}
		&&\multicolumn{1}{c|}{ACE }	&	0.9344& 0.8858&	0.9256&	0.9216	&\multicolumn{1}{c|}{ACE}&0.8687& 0.8293& 0.8627& 0.8599\\
		\cline{2-12}
		\multirow{8}{*}{Ensemble2}
		&\multirow{2}{*}{FGSM}		&$\epsilon$=0.1		& 0.8358&0.8418&0.8316&	\textbf{0.8534} 				&$\epsilon$=0.03& 0.6573& \textbf{0.7886}& 0.6080& 0.6582\\
									&&$\epsilon$=0.3	& 0.6210& \textbf{0.7448}& 0.6970&0.6392 				&$\epsilon$=0.09& 0.5272& \textbf{0.6598}& 0.4838& 0.5256\\
									\cline{2-12}
		&\multirow{2}{*}{MI-FGSM}	&$\epsilon$=0.1		& 0.6268& 	0.6510& 	0.6165& \textbf{0.6866}		&$\epsilon$=0.03& 0.1871& 0.1110& 0.1473& \textbf{0.2061}\\
									&&$\epsilon$=0.3	& 0.4834& 	\textbf{0.4940}& 	0.4571& 0.4326		&$\epsilon$=0.09& 0.0582& 0.0101& 0.0426& \textbf{0.0695}\\
									\cline{2-12}
		&\multirow{2}{*}{PGD}		&$\epsilon$=0.1		& 0.6702& 	0.6605& 	0.6030& \textbf{0.7501}		&$\epsilon$=0.01& 0.5479& \textbf{0.6049}& 0.4913& 0.5759\\	
									&&$\epsilon$=0.15	& 0.5810& 	0.5705&		0.5138&	\textbf{0.6801}		&$\epsilon$=0.02& 0.4027& 0.3867& 0.3383& \textbf{0.4224}\\
									\cline{2-12}
		&\multirow{2}{*}{BIM}		&$\epsilon$=0.1		& 0.6865& 	0.6816& 	0.6366& \textbf{0.7640}		&$\epsilon$=0.01& 0.5929& \textbf{0.6386}& 0.5249& 0.6186\\
									&&$\epsilon$=0.15	& 0.5893& 	0.5908& 	0.5505&	\textbf{0.6902}		&$\epsilon$=0.2& 0.4393& 0.4382& 0.3841& \textbf{0.4603}\\
									\cline{1-12}
		&&\multicolumn{1}{c|}{ACE}	&	0.9339&	0.8082&	0.923&	0.9228	&\multicolumn{1}{c|}{ACE}&0.8744& 0.8286& 0.8604& 0.8532\\
		\cline{2-12}
		\multirow{8}{*}{Ensemble3}
		&\multirow{2}{*}{FGSM}		&$\epsilon$=0.1		& 0.8435&	0.8603&		0.8319&		\textbf{0.8809} 				&$\epsilon$=0.03& 0.6531& \textbf{0.8204}& 0.6090& 0.7535\\
									&&$\epsilon$=0.3	& 0.6292&	\textbf{0.7216}&	0.6821&	0.6395 						&$\epsilon$=0.09& 0.5370& \textbf{0.6778}& 0.4894& 0.6480\\
									\cline{2-12}
		&\multirow{2}{*}{MI-FGSM}	&$\epsilon$=0.1		& 0.6417&	0.6471&	0.6159&	\textbf{0.7225}							&$\epsilon$=0.03& 0.1901& 0.1776& 0.1471& \textbf{0.3088}\\
									&&$\epsilon$=0.3	& \textbf{0.5023}&	0.3957&	0.4596&	0.4531							&$\epsilon$=0.09& 0.0564& 0.0359& 0.0399& \textbf{0.1159}\\
									\cline{2-12}
		&\multirow{2}{*}{PGD}		&$\epsilon$=0.1		& 0.6789&	0.6978&	0.6087&	\textbf{0.7838}							&$\epsilon$=0.01& 0.5489& \textbf{0.6438}& 0.4995& 0.6431\\	
									&&$\epsilon$=0.15	& 0.6085&	0.5673&	0.5126&	\textbf{0.7207}							&$\epsilon$=0.02& 0.4055& 0.4440& 0.3410& \textbf{0.5206}\\
									\cline{2-12}
		&\multirow{2}{*}{BIM}		&$\epsilon$=0.1		& 0.6941&	0.7024&	0.6343&	\textbf{0.7929}							&$\epsilon$=0.01& 0.5933& 0.6751& 0.5253& \textbf{0.6751}\\
									&&$\epsilon$=0.15	& 0.6106&	0.5857&	0.5510&	\textbf{0.7197}							&$\epsilon$=0.2& 0.4486& 0.4930& 0.3916& \textbf{0.5672}\\
									\cline{1-12}
		&&\multicolumn{1}{c|}{ACE}	&	0.9332& 	0.9015& 	0.9244& 	0.9240	&\multicolumn{1}{c|}{ACE}&0.8705&	0.8370&	0.8616&	0.8592\\
		\cline{2-12}
		\multirow{8}{*}{Ensemble4}
		&\multirow{2}{*}{FGSM}		&$\epsilon$=0.1		& 0.8423& 	0.8293& 	0.8370& 	\textbf{0.8631} 				&$\epsilon$=0.03& 0.6464&	\textbf{0.8016}&		0.6001&		0.6646\\
									&&$\epsilon$=0.3	& 0.6255& 	\textbf{0.7114}& 	0.6966& 	0.6582 					&$\epsilon$=0.09& 0.5236&	\textbf{0.7076}&		0.4821&		0.5471\\
									\cline{2-12}
		&\multirow{2}{*}{MI-FGSM}	&$\epsilon$=0.1		& 0.6279& 	0.5801& 	0.6147& 	\textbf{0.6889}					&$\epsilon$=0.03& 0.1754&		0.1144&		0.1555&		\textbf{0.2174}\\
									&&$\epsilon$=0.3	& 0.4813& 	0.3207& 	0.4709& 	\textbf{0.5001}					&$\epsilon$=0.09& 0.0454&		0.0186&		0.0470&		\textbf{0.0690}\\
									\cline{2-12}
		&\multirow{2}{*}{PGD}		&$\epsilon$=0.1		& 0.6677& 	0.6207& 	0.6049& 	\textbf{0.7436}					&$\epsilon$=0.01& 0.5408&		\textbf{0.5887}&		0.4932&		0.5861\\	
									&&$\epsilon$=0.15	& 0.5903& 	0.4892& 	0.5209& 	\textbf{0.6814}					&$\epsilon$=0.02& 0.3885&		0.3778&		0.3398&		\textbf{0.4306}\\
									\cline{2-12}
		&\multirow{2}{*}{BIM}		&$\epsilon$=0.1		& 0.6800& 	0.6495& 	0.6380& 	\textbf{0.7486}					&$\epsilon$=0.01& 0.5808&	\textbf{0.6271}&		0.5218&		0.6228\\
									&&$\epsilon$=0.15	& 0.5970& 	0.5196& 	0.5553& 	\textbf{0.6879}					&$\epsilon$=0.2& 0.4279&		0.4344&		0.3844&		\textbf{0.4696}\\
		\bottomrule
	\end{tabular}}
\end{table*}

\section{Conclusion}
\label{sec: Conclusion}
In this paper, we propose a practical and feasible adversarial examples defense scheme based on negative correlation ensemble, which is named NCEn. 
We use the negative correlation principle to make the gradient direction and gradient magnitude of each member in the ensemble negatively correlated, 
and to train all members in the ensemble interactively and simultaneously.
The purpose of negative correlation training in NCEn is to produce the best defense performance for the whole ensemble.
Experiments have shown that NCEn can reach a better adversarial robustness than other ensemble defense schemes. 
Then, we demonstrate that NCEn can reduce the transferability of adversarial examples between members in the ensemble through the confusion matrix.
Finally, we discuss the influence of hyperparameters $\nabla \lambda_{cos}$ and $\nabla \lambda_{norm}$ on defense performance, 
and give the best settings of hyperparameters.
In general, we can conclude that NCEn is capable to improve the diversity and the robustness of the ensemble by making the gradient direction and gradient magnitude negatively correlated. 
However, for a given task, how to find the optimal number of the models in the ensemble is still a question. 
Many models could not be better. In the future, we will study how to automatically set the optimal number of the models in the ensemble. 

\bibliographystyle{IEEEtran}
\bibliography{ref}

\vspace{-1.05 cm}
\begin{IEEEbiography}[{\includegraphics[width=1in,height=1.25in,clip,keepaspectratio]{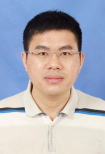}}]{Wenjian Luo}
 	(SM'15, IEEE) received the BS and PhD degrees from Department of Computer Science and Technology, University of Science and Technology of China, Hefei, China, in 1998 and 2003. He is presently a professor of School of Computer Science and Technology, Harbin Institute of Technology, Shenzhen, China. His current research interests include computational intelligence and applications, network security and data privacy, machine learning and data mining. He is a distinguished member of CCF and a senior member of IEEE, ACM and CAAI. He currently serves as an associate editor or editorial board member for several journals including \textit{Information Sciences Journal}, \textit{Swarm and Evolutionary Computation Journal}, \textit{Journal of Information Security and Applications}, \textit{Applied Soft Computing Journal} and \textit{Complex \& Intelligent Systems Journal}. Currently, he also serves as the chair of the IEEE CIS ECTC Task Force on Artificial Immune Systems. He has been a member of the organizational team of more than ten academic conferences, in various functions, such as program chair, symposium chair and publicity chair.
\end{IEEEbiography}

\vspace{-1.05 cm}
\begin{IEEEbiography}[{\includegraphics[width=1in,height=1.25in,clip,keepaspectratio]{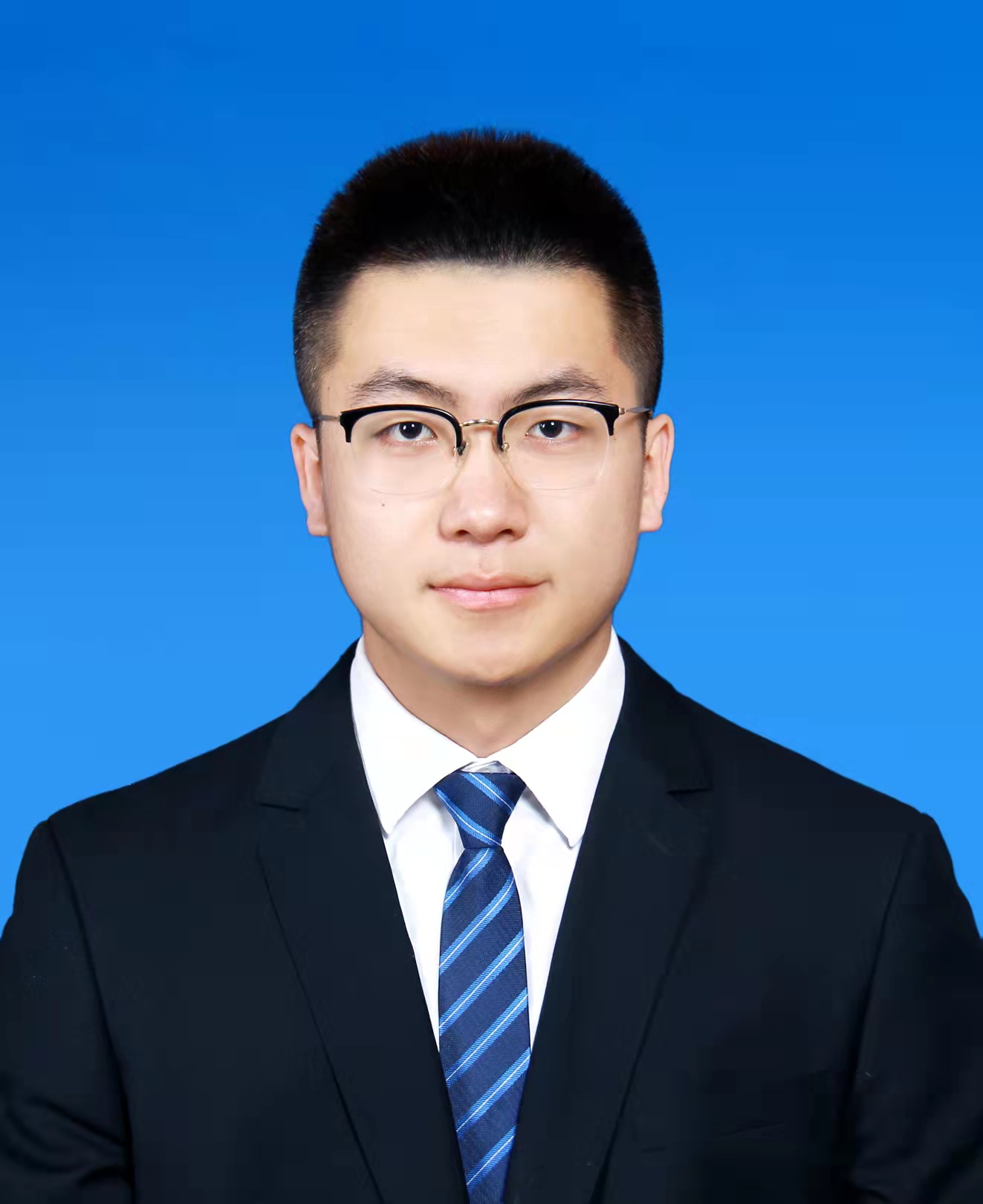}}]{Hongwei Zhang}
	received the B.S. degree from Harbin Institute of Technology, Harbin, China, in 2020. He is currently working toward a Master's degree at the School of Computer Science and Technology, Harbin Institute of Technology, Shenzhen, China. His research interests include adversarial attack and defense.
\end{IEEEbiography}

\vspace{-1.05 cm}
\begin{IEEEbiography}[{\includegraphics[width=1in,height=1.25in,clip,keepaspectratio]{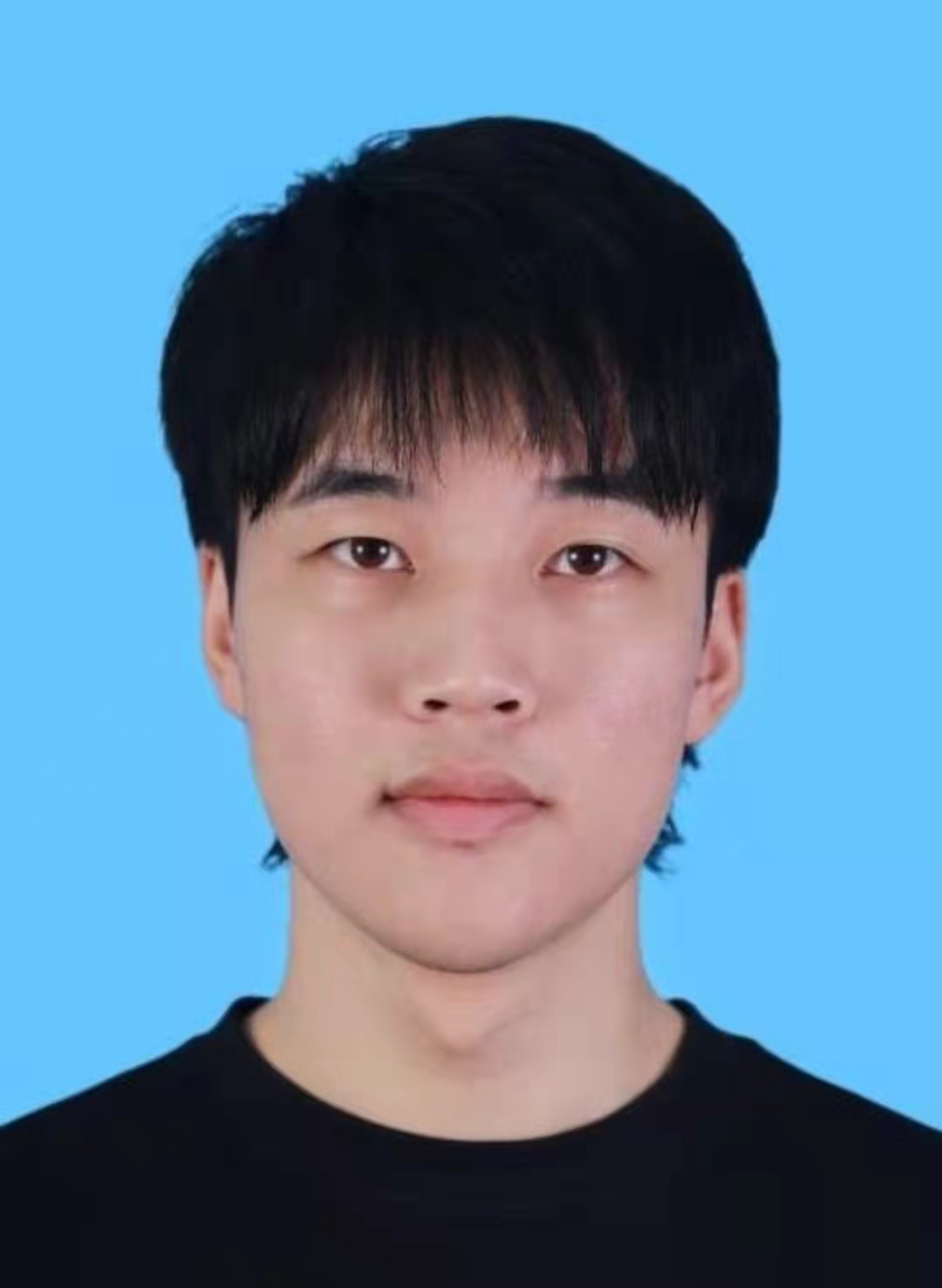}}]{Linghao Kong}
	received the B.E. degree from Northwest University, Shaanxi, China, in 2020. He is currently working toward a Master's degree at the School of Computer Science and Technology, Harbin Institute of Technology, Shenzhen, China. His research interests include adversarial attack and defense.
\end{IEEEbiography}

\vspace{-1.05 cm}
\begin{IEEEbiography}[{\includegraphics[width=1in,height=1.25in,clip,keepaspectratio]{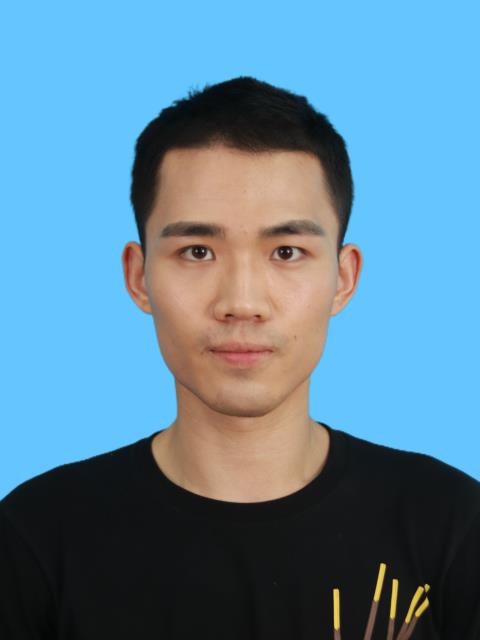}}]{Zhijian Chen}
	received the B.E. degree from Taiyuan University of Technology, Shanxi, China, in 2017. He is currently working towards a Ph.D. degree at the School of Computer Science and Technology, Harbin Institute of Technology, Shenzhen, China. His research interests include machine learning and adversarial examples. 
\end{IEEEbiography}

\vspace{-1.05 cm}
\begin{IEEEbiography}[{\includegraphics[width=1in,height=1.25in,clip,keepaspectratio]{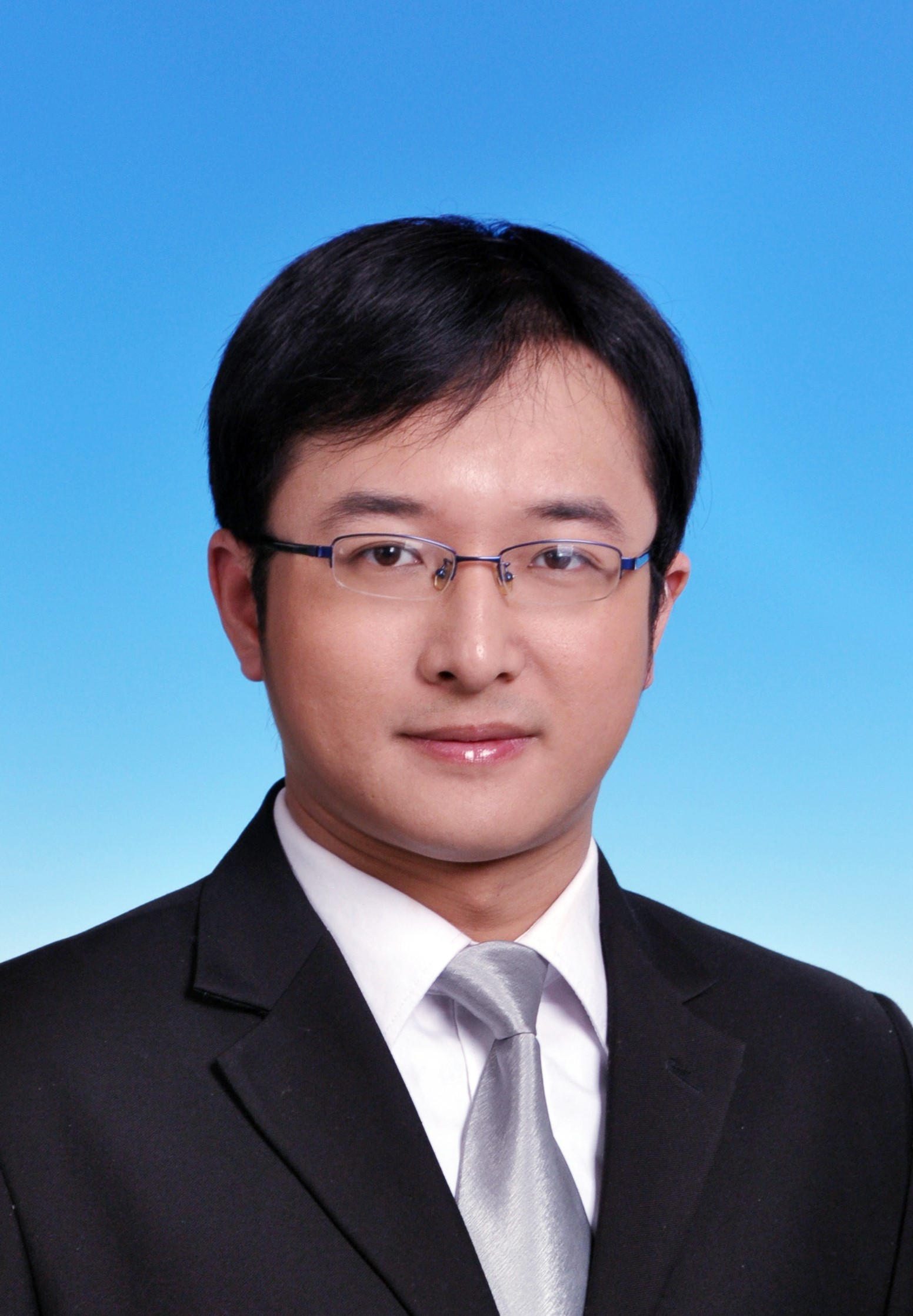}}]{Ke Tang}
	(S'05-M'07-SM'13) received the B.Eng. degree from the Huazhong University of Science and Technology, Wuhan, China, in 2002, and the Ph.D. degree from Nanyang Technological University, Singapore, in 2007. From 2007 to 2017, he was with the School of Computer Science and Technology, University of Science and Technology of China, Hefei, China, first as an Associate Professor from 2007 to 2011 and later a Professor from 2011 to 2017. 
	He is currently a Professor with the Department of Computer Science and Engineering, Southern University of Science and Technology, Shenzhen, China. He has published more than 70 journal papers and more than 80 conference papers. According to Google Scholar, his publications have received more than 8000 citations and the H-index is 42. His major research interests include evolutionary computation, machine learning, and their applications.
\end{IEEEbiography}

\vfill

\end{document}